\documentclass[10pt,twocolumn,aps,prx,superscriptaddress,longbibliography,eprint]{revtex4-2}

\usepackage{graphicx}
\usepackage{amsmath}
\usepackage{amssymb}
\usepackage{float}
\usepackage{braket}   
\usepackage{enumitem}
\usepackage[dvipsnames]{xcolor}
\usepackage[colorlinks=true, urlcolor=blue, linkcolor=blue,citecolor=blue, hypertexnames=false]{hyperref}
\usepackage[detect-weight=true,separate-uncertainty=true]{siunitx}
\usepackage{comment}
\usepackage[normalem]{ulem}

\usepackage{cleveref}
\crefname{equation}{Eq.}{Eqs.}
\Crefname{equation}{Equation}{Equations}
\crefname{figure}{Fig.}{Figs.}
\Crefname{figure}{Figure}{Figures}
\crefname{section}{Sec.}{Sects.}
\Crefname{section}{Section}{Sections}
\crefname{table}{Table}{Tables}
\crefname{appendix}{Appendix}{Apps.}
\Crefname{appendix}{Appendix}{Apps.}

\newcommand{\h}[1]{\hat{#1}}
\newcommand{\ha}{\hat{a}}
\newcommand{\had}{\hat{a}^\dagger}

\newcommand{\hn}{\hat{n}}
\newcommand{\hvarphi}{\hat{\varphi}}

\newcommand{\hH}{\hat{H}}

\newcommand{\zpfr}{\varphi_{{\rm zpfr}}}

\newcommand{\schrodinger}{Schr\"{o}dinger }

\newcommand{\lettersection}[1]{\paragraph*{\color{black}#1.---}}

\begin{document}

\title{Balanced cross-Kerr coupling for superconducting qubit readout}

\author{Alex A. Chapple}
\thanks{These authors contributed equally.}
\affiliation{Institut Quantique and D\'epartement de Physique, Universit\'e de Sherbrooke, Sherbrooke J1K 2R1 Quebec, Canada}

\author{Othmane Benhayoune-Khadraoui}
\thanks{These authors contributed equally.}
\affiliation{Institut Quantique and D\'epartement de Physique, Universit\'e de Sherbrooke, Sherbrooke J1K 2R1 Quebec, Canada}

\author{Simon Richer}
\affiliation{Institut Quantique and D\'epartement de Physique, Universit\'e de Sherbrooke, Sherbrooke J1K 2R1 Quebec, Canada}

\author{Alexandre Blais}
\affiliation{Institut Quantique and D\'epartement de Physique, Universit\'e de Sherbrooke, Sherbrooke J1K 2R1 Quebec, Canada}
\affiliation{CIFAR, Toronto, ON M5G 1M1, Canada}

\date{\today}

\begin{abstract}
Dispersive readout, the standard method for measuring superconducting qubits, is limited by multiphoton qubit-resonator processes arising even at moderate drive powers. These processes degrade performance, causing dispersive readout to lag behind single- and two-qubit gates in both speed and fidelity. In this work, we propose a novel readout method, termed ``\textit{junction readout}''. Junction readout leverages the nonperturbative cross-Kerr interaction resulting from coupling a qubit and a resonator via a Josephson junction. Furthermore, by adding a capacitive coupling in parallel to the junction, Purcell decay induced by the exchange coupling can be suppressed. We also show that junction readout is more robust against deleterious multiphoton processes, and offers greater flexibility for resonator frequency allocation. Crucially, junction readout achieves superior performance compared to dispersive readout while maintaining similar hardware overhead. Numerical simulations show that junction readout can achieve fidelity exceeding $99.99\%$ in under $30$ ns of integration time, making it a promising alternative for superconducting qubit readout with current hardware.
\end{abstract}

\maketitle

\lettersection{Introduction} \label{sec:intro}
Fast and high-fidelity qubit measurement is a cornerstone of quantum information processing and fault-tolerant quantum computing. For example, in quantum error correction (QEC) protocols, each round of QEC relies on rapid single-shot readout of ancilla qubits to detect errors. Recent breakthroughs achieving break-even performance in surface \cite{GoogleQuantumAI:2023,GoogleQuantumAI:2024} and bosonic codes \cite{Ofek:2016,Sivak:2023,Brock:2025} have relied on dispersive readout, a standard tool for measuring superconducting qubits. This readout operates by introducing a qubit-state dependent shift in the resonator frequency, enabling qubit state inference without directly disturbing the qubit \cite{Blais_PRA,Wallraff_disp_readout,RMP}.   

Despite the improvements in readout fidelity and integration times \cite{McClure2016, Walter_disp_readout,Sunada_intrinsic,swiadek2023enhancing,Spring:2024}, dispersive readout still lags behind the performance of the best single- and two-qubit gates. Increasing the readout drive power to enhance fidelity and speed often leads to measurement-induced state transitions \cite{Johnson2012, MIST_1, Lescanne:2019, Dynamics_of_transmon_ionzation, Reminiscence_chaos, MIST_2,hazra_2024, Dumas2024} which are detrimental to error correction protocols as they introduce correlated errors \cite{Varbanov:2020,Miao:2023}. Resetting such states, which typically involve $\sim$ 5 to 10 photons \cite{MIST_2}, is challenging even with the use of leakage reduction techniques~\cite{ McEwen:2021, Battistel:2021, Marques:2023, Lacroix:2023}. Thus, achieving fast, high-fidelity readout with low leakage rates remains an open problem.

\begin{figure}[t]
    \centering
    \includegraphics[width=0.87\linewidth]{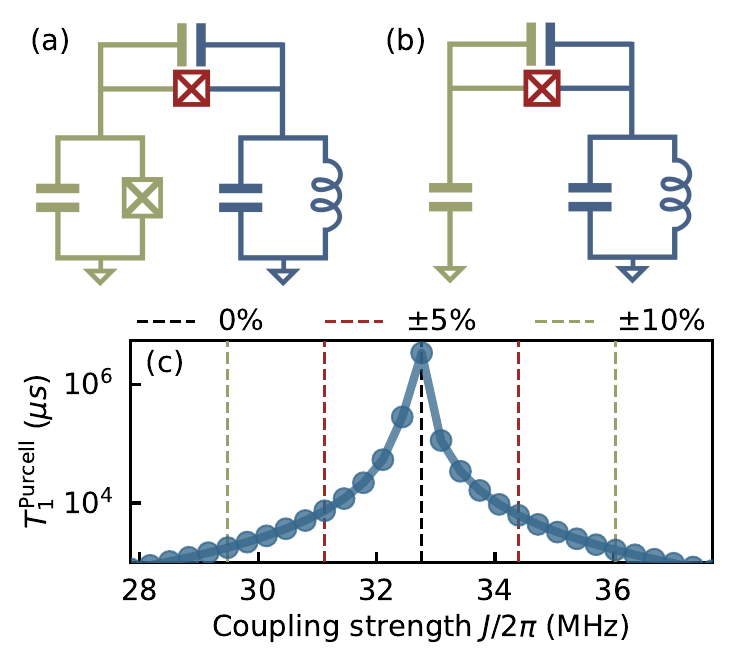}
    \caption{Junction readout circuit (a) with and (b) without flux loop or low-frequency mode. A transmon (green) is coupled to the readout resonator (blue) through a Josephson junction (red) and a capacitor. Unlike in (a), in (b) the transmon nonlinearity is entirely inherited from the coupling junction, see Ref.~\cite{Supplementary_Information}. (c) Jaynes–Cummings induced Purcell lifetime $T_1^{\mathrm{Purcell}}$ of the transmon for varying capacitive coupling strengths $J$.
    The black dashed line indicates where the cancellation condition of \cref{balanced_coupling} is met. There, $J/2 \pi \simeq 32.8 \: \textrm{MHz}$, corresponding to a coupling capacitance around $10 \: \textrm{fF}$. The qubit and resonator frequencies are $\omega_q/2\pi = 5.672$ GHz and $\omega_r/2\pi = 9.375$ GHz, respectively, and the resonator decay rate is $\kappa/2\pi = 8$ MHz, see Sec S2 of \cite{Supplementary_Information}.}
    \label{fig:circuits}
\end{figure}

To further complicate matters, state-of-the-art dispersive readout often requires a Purcell filter to prevent the qubit from decaying through the readout channel~\cite{Reed2010,Jeffrey_purcell}.  While effective, adding a Purcell filter increases the readout system's footprint, complicates calibration, and makes multiplexing more challenging. These challenges have spurred interest in intrinsically Purcell-protected qubits and readout methods as compact, scalable alternatives for next-generation quantum processors~\cite{dimon_Gambetta,  Diniz_CrossKerr, Didier_longitudinal, Remy_cross_kerr, Pmon}.

Here, we propose an approach to mediate a nonperturbative dispersive qubit-resonator interaction, enabling high-fidelity and fast measurement while suppressing the dominant channel contributing to Purcell decay.
Even when accounting for reduced readout efficiency and finite qubit lifetime, our approach achieves an order-of-magnitude improvement over state-of-the-art readout systems with comparable hardware overhead.

\lettersection{Theory of junction readout} \label{sec:theory}
In circuit QED, the dispersive interaction  \( \sum_{i_t} \chi_{i_t} \ket{i_t}\bra{i_t}\hat{a}^\dagger \hat{a} \), with  $\ket{i_t}$ a bare transmon state and $\hat a$ the resonator's annihilation operator, is usually realized by capacitively coupling a qubit and a resonator that are widely detuned in frequency~\cite{RMP}. Here, we seek an alternative circuit that mediates this interaction, with the interaction strength \(2\chi_z = \chi_{1_t} - \chi_{0_t}\) engineered to be nonperturbative (i.e.,~independent of the resonator frequency and not arising from a Schrieffer-Wolff-type frame transformation, in contrast to what one would obtain in the presence of a transverse coupling). 

\begin{figure}
    \centering
    \includegraphics[width=\linewidth]{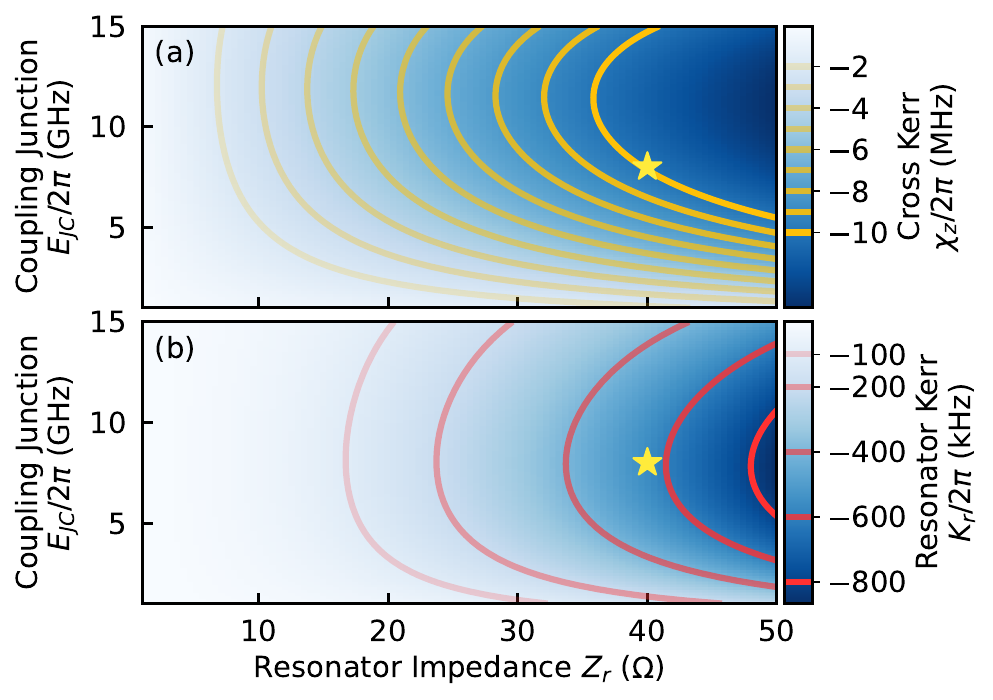}
    \caption{
    (a) Cross-Kerr coupling $\chi_z$ and (b) resonator self-Kerr $K_r$ between the transmon and the resonator for varying resonator impedance $Z_r$ and coupling junction energy $E_{Jc}$. The contour lines indicate lines of constant (a) cross-Kerr ranging from $-2$  to $-10 \: \textrm{MHz}$ and (b) self-Kerr ranging from $-100$ to $-800 \: \textrm{kHz}$. The star marks the parameter used in the readout simulations of \cref{fig:readout_performance}.
    }
    \label{fig:transmon_impedance_coupling}
\end{figure}

Alternative strategies for engineering a nonperturbative Kerr interaction have been explored~\cite{Didier_longitudinal, Diniz_CrossKerr, Remy_cross_kerr, Quarton_ye_2021, ye2024ultrafast, Pmon}. In contrast to these approaches, which often involve complex circuit designs, rely on special symmetries, or encode qubits in spatially delocalized modes, our proposed circuits, shown in \cref{fig:circuits}, offer a simpler solution with hardware overhead comparable to that of dispersive readout. In the circuit \cref{fig:circuits} (a), the transmon (green) is coupled to a readout resonator (blue) through a Josephson junction in parallel with a capacitor whose role will be discussed below. The Hamiltonian of this circuit takes the form
\begin{equation}
\begin{split}
    \hat{H} &= 4 E_C (\hat{n}_t - n_g)^2 - E_J \cos \hat{\varphi}_t \\ 
    &\quad + \omega_r \hat{a}^\dagger \hat{a} \\
    &\quad - E_{J_c} \cos(\hat{\varphi}_t - \hat{\varphi}_r) + J \hat{n}_t\hat{n}_r \\
    &\equiv \hat{H}_{\text{tr}} + \hat{H}_\text{r} + \hat{H}_{\text{int}},
\end{split} \label{eqn:system_hamiltonian}
\end{equation}
where the first (second) line corresponds to the transmon (resonator) Hamiltonian $\hat{H}_{\text{tr}}$ ($\hat{H}_{\text{r}}$) while the last line corresponds to the interaction Hamiltonian $\hat{H}_{\text{int}}$. Furthermore, \(\hat{\varphi}_t\) and \(\hat{n}_t\) are the transmon's phase and charge operators which satisfy the relation $[e^{i\h{\varphi}_t}, \h{n}_t] = -e^{i\h{\varphi}_t}$, and \(E_J\) and \(E_C\) are its Josephson and charging energies, respectively. The resonator has frequency $\omega_r$, and its phase operator is given by $\hat{\varphi}_r= \varphi_\mathrm{zpfr}(\ha+\ha^\dagger)$ where $\zpfr= (2\pi/\Phi_0) \sqrt{\hbar Z_r/2}$ is the phase zero-point fluctuations. Furthermore, $E_{Jc}$ is the Josephson energy of the coupling junction, and $J$ is the capacitive coupling strength between the transmon and the resonator. The gate charge \(n_g\) is included explicitly, as it has been shown to affect the onset of measurement-induced state transitions \cite{Malekakhlagh:2022,Reminiscence_chaos, MIST_2, Dumas2024}. As we show below, our proposed circuit ensures a high measurement critical photon number regardless of the value of the gate charge.

Setting \(J = 0\) for the moment, the coupling Hamiltonian can be written in the form
\begin{equation} \label{coscos_interaction}
\begin{split}
    \hat{H}_{\text{int}} 
    = -E_{J_c} \cos \hat{\varphi}_t \cos \hat{\varphi}_r - E_{J_c} \sin \hat{\varphi}_t \sin \hat{\varphi}_r.
\end{split}
\end{equation}
To second order in phase fluctuations, the cos-cos interaction of the first term leads to \( \hat{\varphi}_t^2 \hat{\varphi}_r^2\), thereby mediating a nonperturbative cross-Kerr coupling. As discussed in Ref.~\cite{chapple:2024}, this results in large measurement critical photon numbers and no Purcell decay. When supplemented with the appropriate resonator drive~\cite{chapple:2024}, this interaction emulates the longitudinal coupling discussed in Ref.~\cite{Didier_longitudinal}, enabling fast and accurate discrimination of the pointer states. Here, we show that the circuit of \cref{fig:circuits} achieves the aforementioned benefits with a significantly simplified design.
On the other hand, to first order in phase fluctuations, the sin-sin interaction term is an unwanted Jaynes-Cummings interaction that causes multiphoton resonances \cite{ MIST_1, Lescanne:2019, Dynamics_of_transmon_ionzation, Reminiscence_chaos, MIST_2},
thereby lowering the critical photon number, see Sec S3A of \cite{Supplementary_Information}. A potential mitigation strategy is to substantially increase the qubit–resonator detuning—well beyond what it is used in standard dispersive readout—while simultaneously increasing the coupling to keep the dispersive shift fixed~\cite{kurilovich:2025, connolly2025, dai2025}. In practice, this places the qubit at a small fraction of the resonator frequency (e.g., $\omega_r/\omega_q \sim 10$) and thus relies on low-frequency qubits strongly coupled to high-frequency resonators, which introduces its own set of challenges.

Instead, here we propose to cancel this unwanted sin-sin term by introducing a parallel capacitance to the Josephson junction which mediates a charge-charge coupling $J \h{n}_t \h{n}_r$, see \cref{fig:circuits}. The coupling strength $J$ is specifically chosen to cancel the exchange (Jaynes–Cummings) coupling by setting the $0 \leftrightarrow 1$ matrix element of the interaction Hamiltonian to zero, as expressed by the condition
\begin{equation} \label{balanced_coupling}
\begin{split}
-&E_{J_c}\bra{1_t,0_r}\sin\hvarphi_t\sin\hvarphi_r\ket{0_t,1_r}\\
&+J\bra{1_t,0_r}\hn_r\hn_t\ket{0_t,1_r}=0,
\end{split}
\end{equation}
where $\ket{j_r}$ denotes a bare resonator state. Intuitively, this condition—which we refer to as balanced cross-Kerr coupling—describes destructive interference between the current paths through the capacitor and the junction that mediate the exchange coupling between the qubit and readout modes; see also Ref.~\cite{kounalakis:2018}. Importantly, since the cancellation condition involves only the computational subspace, the optimal coupling strength \(J\) is insensitive to fluctuations in the gate charge \(n_g\). Moreover, as shown in \cref{fig:circuits}(c), the Purcell decay —associated with the residual exchange coupling— remains minimal even when the condition \cref{balanced_coupling} is not perfectly satisfied. For instance, with a \( \sim 10\%\) imperfection in the junction or capacitor fabrication, the Purcell decay time \(T_1\) remains on the order of \(\sim 1\ \text{ms}\) for the chosen parameters \cite{Supplementary_Information}, highlighting the robustness of this scheme to fabrication errors. In Sec S8, we analyze the circuit shown in \cref{fig:circuits}(b), which also realizes  \cref{eqn:system_hamiltonian} but without involving a flux loop, along with other alternative circuit implementations.

\lettersection{Choice of readout parameters} \label{sec:chivskerr}
We now turn to a discussion of the optimal choice of parameters. To achieve fast readout we aim for a large transmon-resonator cross-Kerr interaction. However, as is evident from \cref{coscos_interaction}, a nonperturbative cross-Kerr term (resulting from \( \hat{\varphi}_t^2 \hat{\varphi}_r^2\)) is inherently accompanied by a nonperturbative self-Kerr nonlinearity on the resonator (resulting from \(\hat{\varphi}_r^4\)). Resonator self-Kerr can distort the coherent state in the resonator~\cite{Sivak2019,Boutin2017}, potentially reducing readout fidelity by hindering the clear separation of pointer states and limiting the maximum photon population in the resonator. Moreover, this distortion renders the conventional linear measurement filter suboptimal for state assignment \cite{Bultink2018,Gembetta_optimal_readout, Danjou:2014}. 

To achieve a large cross-Kerr interaction ($|\chi_z| / 2\pi \sim 2 - 10$ MHz) without compromising the linearity of the readout resonator, we optimize the resonator impedance $Z_r$. From \cref{coscos_interaction}, the leading contribution of the cross-Kerr strength is given by 
$ \chi_z \simeq -\varphi_\mathrm{zpfr}^2 E_{Jc} \sqrt{2 E_C/E_{J, \rm{total}}} /2$, with $E_{J, \rm{total}} = E_J + E_{Jc}$ which we keep constant. The self-Kerr nonlinearity of the resonator inherited from the coupling junction is approximately $ K_r \simeq -E_{Jc}\varphi_\mathrm{zpfr}^4/2.$ As a result, the cross-Kerr interaction decreases quadratically with the resonator phase zero-point fluctuations $\zpfr$, while the self-Kerr decreases quartically with $\zpfr$. Conversely, both $\chi_z$ and $K_r$ increase linearly with respect to the coupling junction energy $E_{Jc}$. Since $\zpfr$ depends on the resonator impedance \(Z_r\), we optimize \(Z_r\) and \(E_{Jc}\) to achieve a large cross-Kerr coupling with minimal self-Kerr nonlinearity ($|K_r| / 2\pi \lesssim 500$ kHz). \cref{fig:transmon_impedance_coupling}(a) and (b) show the cross-Kerr coupling strength \(\chi_z\) and the resonator's self-Kerr \(K_r\) extracted from exact numerical diagonalization of \cref{eqn:system_hamiltonian} for a range of resonator impedances $Z_r$ and coupling strengths \(E_{Jc}\). 
For realistic circuit parameters, we are able to achieve a cross-Kerr coupling comparable to standard dispersive readout, typically ranging from 2 to 10 MHz. Furthermore, by slightly reducing the resonator impedance below $50~\Omega$---something that is easily achievable experimentally---the self-Kerr nonlinearity can be tuned to match, or in some cases even fall below, the typical values of 100 to 500 kHz observed in standard dispersive readout. 

Interestingly, we also observe that, in junction readout, the dispersive shift shows minimal variation with increasing resonator photon numbers, in contrast to dispersive readout, where the dispersive shift decreases significantly. This difference arises from the nature of the  qubit-resonator interaction which is different in the two schemes. Indeed, in junction readout, the cos-cos interaction leads to smaller higher-order corrections to the dispersive shift at large photon numbers compared to dispersive readout, enabling a faster readout as the cross-Kerr coupling remains large even when the resonator is populated, see \cite{Supplementary_Information} for further details. 
 
The above analysis reveals that junction readout produces qubit-state-dependent frequency shifts similar to those of dispersive readout, suggesting comparable performance at first glance. In the following, we highlight the significant advantages offered by junction readout, demonstrating its superiority over dispersive readout in key aspects, namely its larger robustness against measurement-induced state transitions, as well as faster and higher fidelity measurements.

\lettersection{Suppressing ionization}
The quantum nondemolition (QND) nature of dispersive readout is challenged by multiphoton processes arising due to accidental degeneracies between the qubit and the resonator~\cite{MIST_1}. This phenomenon, also referred to as ionization, occurs when two states of the transmon-resonator system become resonant as photons populate the resonator, leading to a sudden population transfer from the resonator to the transmon ~\cite{MIST_1, MIST_2, Lescanne:2019,Reminiscence_chaos, Dynamics_of_transmon_ionzation, Dumas2024, Xiao2023}. Because this process can typically involve highly excited states of the transmon which are charge sensitive, the resonator critical photon number $n_{\rm{crit}}$  at which these multiphoton processes occur can fluctuate widely with gate charge~\cite{MIST_2, Dumas2024, Reminiscence_chaos, Malekakhlagh:2022}. Note that $n_{\rm{crit}}$ here differs from the Jaynes-Cummings $n_{\rm{crit}}^{jc} = ( \Delta / 2 g)^2$ which marks the breakdown of the dispersive approximation in dispersive readout \cite{RMP}. We now show junction readout is far more robust against such multiphoton processes than dispersive readout.

\begin{figure}[t]
    \centering
    \includegraphics[width=\linewidth]{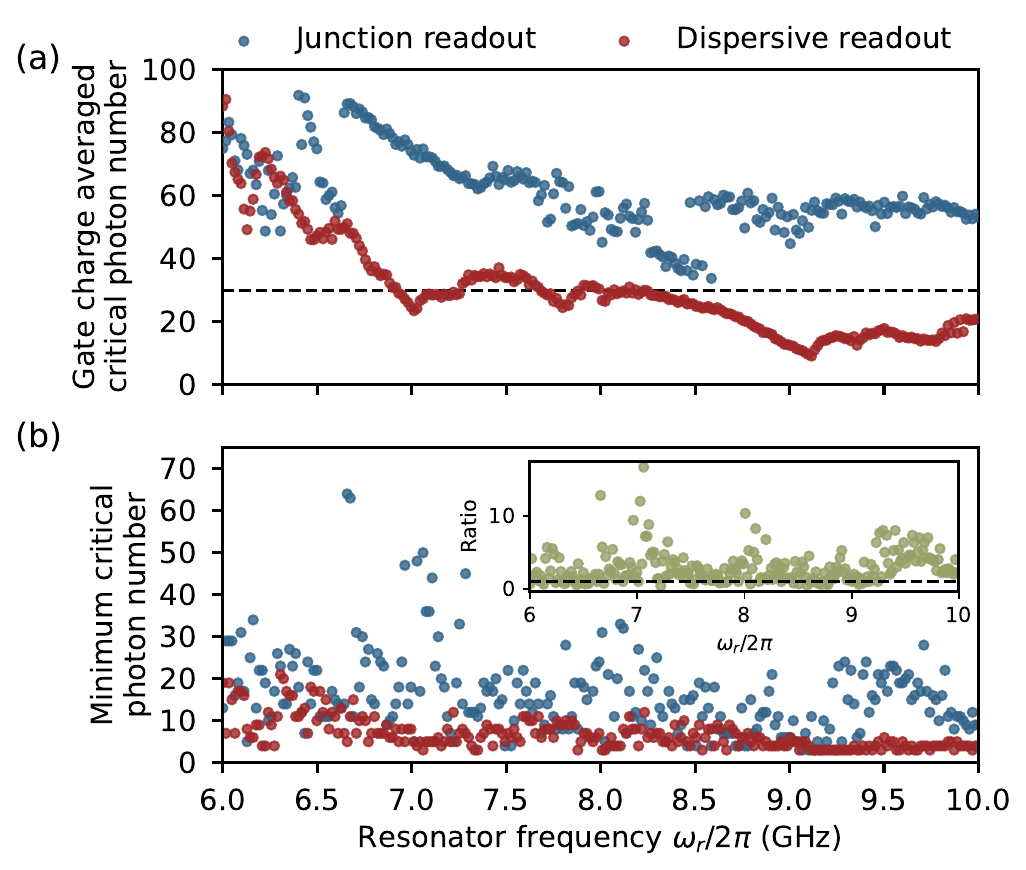}
    \caption{(a) Critical photon number averaged over gate charge $n_g \in [0,0.5]$ as a function of resonator frequency for junction readout (blue) and dispersive readout (red). In both cases, the dispersive shift is fixed at $|\chi_z|/2\pi \simeq 9$ MHz. Dashed horizontal line indicates $n_{\rm{crit}} = 30$. (b) Minimum critical photon number over $n_g \in [0,0.5]$. Inset: ratio of the minimum critical photon numbers between junction and dispersive readout. Dashed horizontal line indicates where the ratio is $1$.}
    \label{fig:transmon_omr_ng_optimized}
\end{figure}

\begin{figure*}[t]
    \centering
    \includegraphics[width=\linewidth]{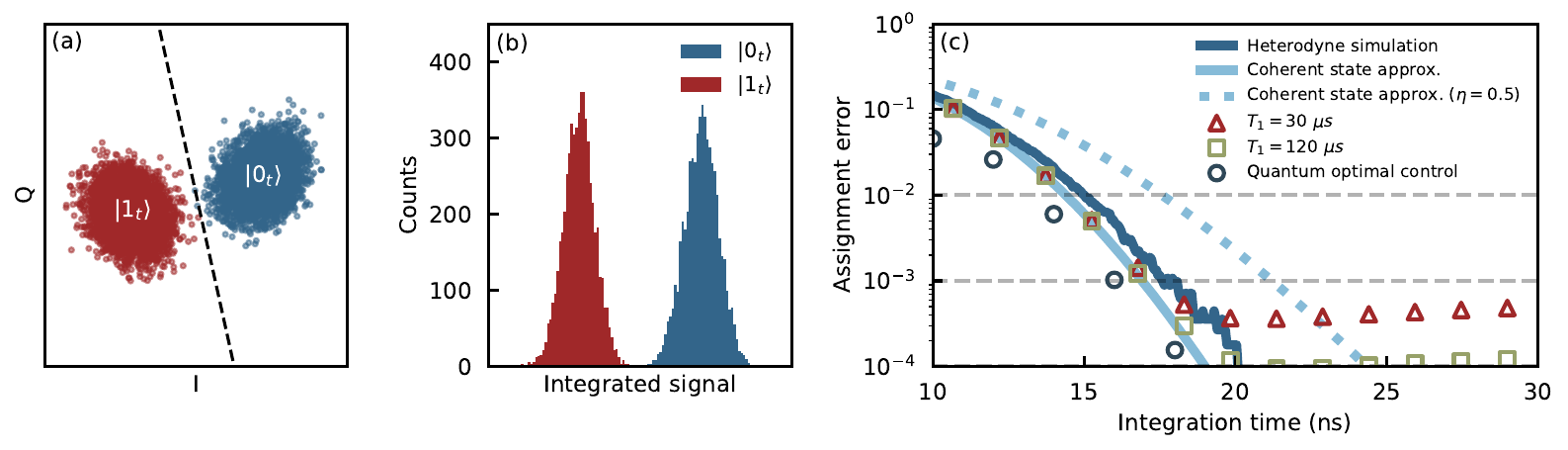}
    \caption{(a) IQ-plane of 11,200 single-shot heterodyne readout simulations of a transmon with junction readout. The black dashed line is the optimal discriminator of the two blobs corresponding to the qubit prepared in the ground (blue) or excited state (red). (b) Histogram of the integrated signal using an optimal discriminator. For both (a) and (b) the integrated time is $t_m \simeq 20 \: \textrm{ns}$. (c) Assignment error obtained from the stochastic heterodyne readout simulations compared to coherent state approximated assignment error. Using the coherent state approximated assignment error, we also show the assignment error for readout efficiency of $\eta = 0.5$ (where $0 \leq \eta \leq 1$) as well as when $T_1$ is $30$ or $120 \: \mu s$. Moreover, we show that quantum optimal control (QOC) further improves the readout fidelity. Here, the transmon charging energy is $E_C / 2 \pi = 300 \: \textrm{MHz}$ with $E_{J, \rm{total}}/E_C = 50$ and gate charge $n_g = 0.0$. The resonator frequency is $\omega_r / 2 \pi = 9.375 \: \textrm{GHz}$, with resonator impedance $Z_r = 40 \: \Omega$ and decay rate $\kappa/2\pi = 2 |\chi_z| \simeq 20$ MHz. The junction coupling strength is $E_{Jc} / 2 \pi = 8 \: \textrm{GHz}$. This set of parameter results in a cross-Kerr of $|\chi_z| / 2\pi \simeq 10 \: \textrm{MHz}$, and a critical photon number of $n_{\rm{crit}} = 65$.
    }
    \label{fig:readout_performance}
\end{figure*}

To assess this robustness, we compute the critical photon numbers using branch analysis, a numerical tool predicting the onset of ionization and that has been shown to match experimental observations \cite{Dumas2024}; see Ref.~\cite{Supplementary_Information} for further details in the context of junction readout. \cref{fig:transmon_omr_ng_optimized}(a) shows the critical photon number averaged over different realizations of the gate charge $n_g \in [0,0.5]$, while panel (b) shows the minimum critical photon number across those realizations, for both junction readout (blue) and dispersive readout (red), over a wide range of resonator frequencies $\omega_r / 2\pi \sim 6 - 10 \: \textrm{GHz}$. To ensure a fair comparison between both approaches, the dispersive shift is fixed to $|\chi_z| / 2\pi \simeq 9 \: \textrm{MHz}$ across all values of $\omega_r$. For the dispersive readout, this requires adjusting the qubit-resonator coupling at all values of the resonator frequency~\cite{RMP}. 
We observe that junction readout consistently yields a higher average critical photon number than dispersive readout over nearly the entire range of resonator frequencies considered. Moreover, we show in Ref.~\cite{Supplementary_Information} that for junction readout, there exists a broad region of resonator frequencies in which over $95\%$ of gate charge realizations yield a critical photon number exceeding 30 (dashed horizontal line in \cref{fig:transmon_omr_ng_optimized}(a)). In contrast, no such frequency range exists for dispersive readout where a comparable fraction of realizations consistently surpass this threshold—see \cite{Supplementary_Information} for further details. Crucially, even in the worst-case scenario, the minimum critical photon number over realizations of the gate charge is larger for junction readout than for dispersive readout in most cases, as shown in \cref{fig:transmon_omr_ng_optimized}(b). 

The markedly higher ionization critical photon numbers observed in junction readout can be largely attributed to the cancellation condition of \cref{balanced_coupling}. Indeed, from perturbation theory, when the transmon ionizes to an excited state outside of the computational manifold, the transition occurs through virtual excitations, sequentially climbing the intermediate states between the initial state and the final ionized state~\cite{Dumas2024}. The cancellation condition in \cref{balanced_coupling}, however, eliminates the matrix element responsible for the transition between the ground state and the first excited state, thereby preventing the initial step of leaving the computational subspace from the ground state. Additionally, while not exact, this cancellation approximately suppresses the matrix element responsible for transitions from the first to the second excited state, further reducing ionization when the qubit begins in its first excited state, see Sec S3A of \cite{Supplementary_Information} for more details.

Finally, we further note that dispersive readout has less flexibility for frequency allocation as the dispersive shift depends on the resonator frequency, and often requires the readout resonator to be close in frequency to the qubit for a large dispersive shift. On the other hand, junction readout relies on a nonperturbative cross-Kerr interaction which does not depend on the resonator frequency, and thus offers greater flexibility for frequency allocation and potentially can alleviate frequency crowding issues for large-scale chips, whilst maintaining a large cross-Kerr for fast readout. 

\lettersection{Readout performance} We have shown that junction readout can realize large cross-Kerr couplings together with small resonator self-Kerrs, higher critical photon numbers compared to dispersive readout across a wide range of resonator frequencies, and extended Purcell lifetimes. The ability to drive the resonator to large populations without encountering multiphoton resonances or significant leakage is crucial for achieving high-fidelity readout. In the following, we present numerical simulations of junction readout suggesting that fast, high-fidelity, and quantum nondemolition readout of transmons can be achieved with realistic parameters.

Our simulations are based on integrating the stochastic \schrodinger equation for heterodyne measurements of the resonator first assuming an ideal readout efficiency $\eta =1$~\cite{Wiseman1993}, see \cite{Supplementary_Information} for further details. Because they use the full Hamiltonian \cref{eqn:system_hamiltonian}, these simulations account for 
the cross-Kerr, $|\chi_z| / 2 \pi \simeq 10 \: \textrm{MHz}$, but also the deleterious effect of the resonator self-Kerr, $|K_r| / 2 \pi \simeq 489 \: \textrm{kHz}$. A two-step measurement pulse leading to an average photon number of $\Bar{n} \simeq 40$ photons is used, well below the critical photon number of $n_{\rm{crit}} = 65$. Despite this short integration time, we find a large separation of the measurement result distribution for the two initial state using an optimal linear discriminator (black dashed line), see \cref{fig:readout_performance}(b). The resulting assignment error vs the integration time $t_m$ (full dark blue line) is reported in panel (c). For $\eta = 1$, a measurement fidelity of 99.99\% is obtained in as short as $t_m = 20 \: \textrm{ns}$. Furthermore, we find a QNDness of $99.89\%$ and $99.71\%$ for the ground and excited state, respectively \cite{Supplementary_Information}.

Because of the small resonator self-Kerr, the distortion of the qubit-state-dependent resonator coherent states is relatively small, see \cref{fig:readout_performance}(a). Under this condition, the resonator state can be approximated as coherent, allowing us to compare our numerical results with the analytical expression for the signal-to-noise ratio (SNR) and the assignment error~\cite{Supplementary_Information, Gambetta:2007,RMP}. The assignment error derived from this approximation (full light blue line) closely matches the full stochastic simulations. The small discrepancy is primarily due to the self-Kerr nonlinearity of the resonator. Given the good agreement between the numerical simulation and the coherent-state approximation, we use the latter to estimate the readout fidelity for finite transmon lifetime and nonideal readout efficiency, something which would be otherwise numerically challenging. First, for $T_1 = 30\:\mu s$ (red triangles), the readout fidelity comfortably exceeds $99.9\%$ in $t_m \simeq 17 \: \textrm{ns}$. For a longer-lived transmon with $T_1 = 120\: \mu s$ (green squares), the fidelity surpasses $99.99\%$ at $t_m \simeq 21 \: \textrm{ns}$. Secondly, assuming an infinite $T_1$ but a  readout efficiency of $\eta = 0.5$ consistent with state-of-the-art experiments \cite{Spring:2024, hazra_2024, Walter_disp_readout}, we find a similar readout performance (dashed light blue line). Using quantum optimal control (QOC), the assignment error can be further reduced, enabling a readout fidelity of $99.99\%$ within $t_m \simeq 18$ ns using realistic pulse shapes. With QOC, the same fidelity can also be achieved in $75$ ns while including resonator reset \cite{Supplementary_Information}. We emphasize that these results are based on parameters readily achievable with current hardware. Even accounting for reduced efficiency and $T_1$ limitations, our results indicate that junction readout could achieve a fidelity exceeding $99.99\%$ in under $30\ \text{ns}$ of integration time, outperforming state-of-the-art readout experiments by an order of magnitude. 

\lettersection{Conclusion}
We have presented a circuit mediating a nonperturbative cross-Kerr coupling between a qubit and a readout resonator. The proposed junction readout method outperforms state-of-the-art dispersive readout by achieving higher critical photon numbers across a broad range of resonator frequencies and is less susceptible to the effect of gate charge. Notably, junction readout enhances the Purcell lifetime and has comparable hardware overhead to dispersive readout. Even when accounting for finite qubit lifetimes and reduced readout efficiency, this approach achieves a readout fidelity of $99.99\%$ in under $30$ ns of integration time. Junction readout overcomes many of the limitations posed by dispersive readout and can be implemented readily with only small modifications to current hardware. Junction readout could become a standard method for superconducting qubit measurement in next-generation quantum processors.

\lettersection{Note}
During the preparation of this manuscript, we learned of similar work experimentally demonstrating Josephson junction-based readout of a transmon qubit~\cite{wang:2024}. In contrast to our approach, in that work the resonator nonlinearity induced by the coupling junction is large such that a high-power bifurcation readout is used.

\lettersection{Acknowledgments}
The authors are grateful to Crist\'obal Lled\'o for a critical reading of the manuscript, and to Peter Spring and Genevi\`eve Gervais for insightful discussions. This work was undertaken thanks in part to funding from the U.S. Army Research Office Grant No. W911NF-23-1-0101. Additional support is acknowledged from NSERC, the Canada First Research Excellence Fund, and the Ministère de l’Économie et de l’Innovation du Québec. 

\let\oldaddcontentsline\addcontentsline
\renewcommand{\addcontentsline}[3]{}
\bibliography{articles}
\let\addcontentsline\oldaddcontentsline

\onecolumngrid
{\center \bf \large Supplemental Material for ``Balanced Cross-Kerr Coupling for Superconducting Qubit Readout''\\}
{\center Alex A. Chapple,\textsuperscript{1,*} Othmane Benhayoune-Khadraoui,\textsuperscript{1,*}, Simon Richer,\textsuperscript{1} and Alexandre Blais\textsuperscript{1,\,2}\\\vspace*{-0.2cm}}
{\center \small \textsuperscript{1}\textit{Institut Quantique and D\'epartement de Physique,\\Universit\'e de Sherbrooke, Sherbrooke J1K 2R1 Quebec, Canada}\\\vspace*{-0.3cm}}
{\center \small \textsuperscript{2}\textit{Canadian Institute for Advanced Research, Toronto, ON, Canada}\\\vspace*{1.cm}}
\twocolumngrid

\setcounter{equation}{0}
\setcounter{figure}{0}
\setcounter{table}{0}
\setcounter{page}{1}
\setcounter{section}{0}

\makeatletter
\renewcommand{\theequation}{S\arabic{equation}}
\renewcommand{\thefigure}{S\arabic{figure}}
\renewcommand{\thesection}{S\arabic{section}}
\renewcommand{\thetable}{S\arabic{table}}

\section{Basics of branch analysis} \label{Branch_analysis}

We outline the basic theory behind branch analysis, a numerical tool used to predict the critical photon number for the onset of ionization. For a more detailed discussion of this approach, see Refs.~\cite{Boissonneault_qubit_readout,Dynamics_of_transmon_ionzation, Dumas2024}. 

To simplify the discussion in this section, we consider the transmon qubit, however, the method also works for other types of qubits. We begin with the undriven, coupled transmon (tr) and resonator (r) Hamiltonian generically given by 
\begin{align}
    \h{H} = \hat{H}_{\rm{tr}} + \hat{H}_{\rm{r}} + \h{H}_{\rm{int}}.
\end{align}
Diagonalizing the Hamiltonian results in a set of eigenvectors $\{ \vert \lambda \rangle \}$. For each considered level $j_t$ of the transmon, we find the eigenvector from $\{ \vert \lambda \rangle \}$ that maximizes the overlap $\vert \langle j_t, 0_r \vert \lambda \rangle \vert^2$ and label this eigenvector as $\vert \overline{j_t, 0_r} \rangle$, the overline denoting a dressed state. We then use these states $|\overline{j_t, 0_r}\rangle$ to recursively assign a label to all remaining states: $\ket{\overline{j_t, n_r + 1}}$ is defined as the unassigned eigenstate $|\lambda \rangle$ which maximizes the overlap $|\langle \lambda | \hat{a}^\dagger | \overline{j_t, n_r} \rangle|^2$. Repeating this for each transmon index $j_t$ results in a set of dressed states $B_{j_t} = \{ \ket{\overline{j_t, n_r}} \}$ which we refer to as branches.

Using these branches, we can compute the average transmon and resonator populations. An ionization process can be identified when the branches for the ground or excited state swaps with a branch of a higher excited state. We define the critical photon number for the ground (excited) state to be when the transmon population exceeds $N_t = 2 \: (3)$ for at least 3 photons in a row, see Refs.~\cite{Dumas2024, chapple:2024} for further details. The definition of the ionization critical photon number used here was shown to agree well with experimental results \cite{Dumas2024}.

\section{Extracting Jaynes–Cummings induced Purcell decay} \label{sec:extract_purcell}
In this section, we discuss how the Purcell-limited lifetimes $T_1^{\rm{Purcell}}$ shown in Fig.~1(c) are computed. For each value of the charge-charge coupling \(J\), we simulate a \(T_1\) experiment by initializing the qubit in its dressed excited state \(\ket{\overline{1_t, 0_r}}\) and evolving it according to 
\begin{equation}
    \frac{d\hat{\rho}}{dt}= -i[\hat{H},\hat{\rho}] + \kappa \mathcal{D}[\hat{a}']\hat{\rho},
\end{equation}
where \(\hat{H}\) is given by Eq.~(1). The transmon parameters are set to \(E_C/2\pi = 300 \:\mathrm{MHz}\), \(E_{J, \rm{total}}/E_C = 50\), and the resonator parameters to \(\omega_r/2\pi = 9.375 \:\mathrm{GHz}\), \(Z_r = 25 \:\Omega\), and \(\kappa/2\pi \simeq 2|\chi_z|/2\pi = 8 \:\mathrm{MHz}\). The coupling Josephson junction is given by \(E_{J_{c}}/2\pi = 4 \:\mathrm{GHz}\).

In the above expression, the collapse operator \(\hat{a}'\) is expressed in the dressed basis as~\cite{BRE:2002}
\begin{equation} \label{cop_purcell}
    \hat{a}' = \sum_{\substack{\lambda',\lambda\\E_{\lambda'}\geq E_\lambda}} \bra{\lambda} \hat{a} \ket{\lambda'} \ket{\lambda}\bra{\lambda'},
\end{equation}
where \(\{\ket{\lambda}\}\) are the eigenstates of the transmon-resonator Hamiltonian \(\hH\) as defined above. Here, we consider only energy-loss processes. Note that with this standard definition of the collapse operator, the resulting relaxation time $T_1^\text{Purcell}$ arises solely from qubit–resonator hybridization due to their exchange (Jaynes–Cummings) coupling. To extract \(T_1^\mathrm{Purcell}\), we fit the excited-state population \(\ket{\overline{1_t, 0_r}}\bra{\overline{1_t, 0_r}}\) to a single exponential decay, \(e^{-t/T_1^{\mathrm{Purcell}}}\). 

The Purcell decay obtained from this method assumes a flat bath spectrum, i.e., \(\kappa(\omega) = \kappa(\omega_r) \equiv \kappa\). In practice, however, the qubit probes the bath at its own frequency, which for these parameters is \(\overline{\omega}_q/2\pi = 5.672\) GHz, rather than the resonator frequency. Therefore, to achieve a more accurate estimate, we normalize the $T_1^{\rm{Purcell}}$ value obtained above by a factor of \(\kappa / \kappa(\overline{\omega}_q)\) ~\cite{Boissonneault_dressed_dephasing, RMP, Labarca2024}, which is approximately \((\overline{\omega}_r / \overline{\omega}_q)^2 \simeq 2.75\), assuming that all photon losses originate from coupling to the readout feedline~\cite{RMP, Labarca2024}. Finally, since the Jaynes–Cummings–induced Purcell decay $T_1^{\text{Purcell}}$ scales as $\sim \kappa \times(g/\Delta)^2$, with $g$ the effective Jaynes–Cummings coupling that itself scales linearly with $J$ as well as with the circuit parameters $E_{J_c}$ and $\sqrt{\omega_q}$ \cite{RMP}, varying these parameters leads to the same conclusion.

\section{Transmon ionization with junction readout}

In this section, we outline how the cancellation condition in junction readout suppresses measurement-induced transitions out of the computational subspace, and discuss physically meaningful metrics for relating critical photon numbers to readout performance. 

\subsection{Effect of cancellation condition on ionization} \label{sec:cancel_ionization}

\begin{figure}
    \centering
    \includegraphics[width=\linewidth]{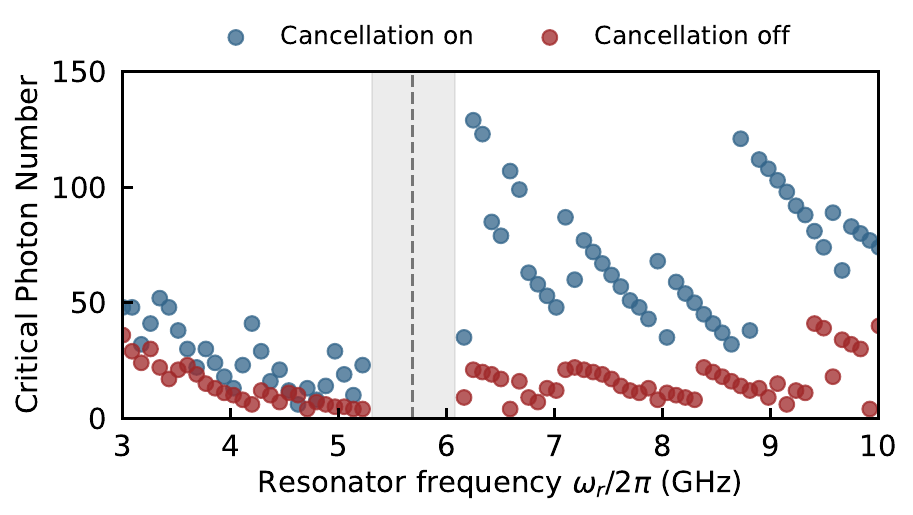}
    \caption{Critical photon numbers for varying resonator frequencies $\omega_r$. The red dots indicate when there is no coupling capacitance, and the blue dots are for where the cancellation condition is met. Here the transmon parameters are the same as in Fig.~4 and the resonator impedance is $Z_r = 35 \: \Omega$. The coupling junction energy is $E_{Jc} / 2\pi = 7.8$ GHz, resulting in $|\chi_z|/2 \pi \simeq 9$ MHz when the cancellation condition is met.}
    \label{fig:ncrit_cancellation}
\end{figure}

As stated in the main text, the matrix element eliminated by the cancellation condition Eq.~(3) is also precisely the matrix element responsible for transitioning the ground state of the transmon to the first excited state when measurement-induced transitions to higher excited states occur. Additionally, the matrix element that connects the first excited state to the second excited state is also approximately suppressed when the condition Eq.~(3) is satisfied. Therefore, by eliminating these matrix elements the likelihood of multiphoton processes causing the transmon to jump in excitation is expected to be reduced. 

\cref{fig:ncrit_cancellation} shows the critical photon numbers for varying resonator frequencies, comparing when the cancellation condition Eq.~(3) is satisfied with the no-cancellation case ($J=0$). When the cancellation condition is met, we observe as expected that the critical photon numbers increase. We also find that canceling the $1\leftrightarrow2$ matrix element, rather than the $0\leftrightarrow1$ element as specified in Eq.~(3), yields essentially the same result: the optimal coupling $J$ and the corresponding critical photon numbers remain essentially unchanged.

It is worth mentioning that for the dispersive readout,  canceling the $0 \leftrightarrow 1$ matrix element would also result in canceling most of the charge-charge coupling-based dispersive shift that is essential for readout. Alternatively, as recently discussed in Ref.~\cite{Sank:2024} in the context of dispersive readout, one could cancel parasitic matrix elements involving higher-energy states that are causing ionization without compromising the dispersive shift. However, no parameter set can cancel \textit{all} parasitic matrix elements simultaneously. Furthermore, since these cancellations involve higher-energy states, they are sensitive to fluctuations in the gate charge. Consequently, this balanced coupling strategy does not push ionization to larger photon numbers for dispersive readout.  

\subsection{Critical photon number percentile}\label{sec:ncrit_threshold}

\begin{figure}
    \centering
    \includegraphics[width=\linewidth]{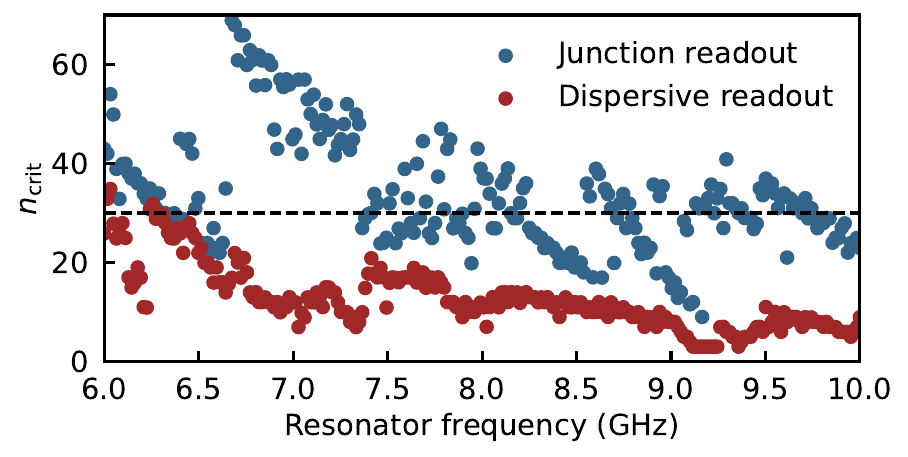}
    \caption{
    Critical photon number at the 5th percentile across resonator frequencies, defined as the value of $n_{\rm crit}$ that $95\%$ of realizations exceed, for both junction readout (blue) and dispersive readout (red). Transmon parameters are the same as in Fig.~(3). Dashed horizontal line indicates $n_{\rm{crit}} = 30$.}
    \label{fig:transmon_omr_ng_percentage_optimized}
\end{figure}

We further compare dispersive and junction readout in terms of their critical photon numbers, and discuss how a larger critical photon number enables higher-fidelity readout. \cref{fig:transmon_omr_ng_percentage_optimized} shows the $5$th percentile critical photon number for each resonator frequency, defined as the value of $n_{\rm crit}$ that $95\%$ of realizations exceed, for both junction readout (blue) and dispersive readout (red). For all data points and both readout schemes, the dispersive shift is fixed at $|\chi_z| / 2\pi \simeq 9$ MHz. As noted in the main text, junction readout exhibits a broad range of resonator frequencies where $95\%$ of the critical photon numbers exceed $30$ for random transmon gate charges, whereas no such range exists for dispersive readout. With this threshold of $n_{\rm crit} = 30$ and the parameters $\chi_z$ and $K_r$ reported in the main text, high performance readout can be achieved.

To see that, we begin with the analytical expression for the signal-to-noise ratio (\rm{SNR}) in the long-time limit \cite{Gambetta2008},
\begin{align} 
    \rm{SNR} \simeq \frac{2 \epsilon}{\kappa} \sqrt{2 \eta \kappa t_m}. 
\end{align}
In the above expression, $\epsilon$ is the drive amplitude, $\kappa$ is the resonator decay rate, $\eta$ is the measurement efficiency, and $t_\mathrm{m}$ is the measurement time. Assuming that the resonator is driven at its Lamb-shifted frequency, and for the optimal ratio $\kappa \simeq 2 |\chi_z|$, which we ensure to be approximately satisfied in our numerical simulations, we can relate the above expression of $\rm{SNR}$ to the average photon number in the resonator $\Bar{n}$ as
\begin{align} 
    \mathrm{SNR} \simeq \sqrt{4 \eta \Bar{n} \kappa t_\mathrm{m}}, 
\end{align}
which simplifies to $\mathrm{SNR} \simeq 2 \sqrt{2 \eta \Bar{n}}$ for a measurement time of order $t_\mathrm{m} \sim 1 / \vert \chi_z \vert$, corresponding to the resonator reaching its steady-state \cite{Walter_disp_readout}.

Choosing the threshold to be $n_{\rm{crit}} = 30$, we must ensure that the maximum number of photons placed in the resonator remains well below this threshold to minimize leakage. 
Given the approximately coherent-state nature of the resonator field during readout, as discussed in the main text, we model the photon number distribution as Poissonian and require its tail to have minimal overlap with the critical photon number. For instance, a safe metric is to limit this overlap to be approximately $0.01\%$. This condition translates to an average photon number $\Bar n$ satisfying 
 $n_{\rm{crit}} = \Bar{n} + 4 \sqrt{\Bar{n}}$, which yields $\Bar{n} \simeq 14.7$. 

For a measurement efficiency of $\eta = 0.5$, in line with state-of-the-art experiments \cite{Spring:2024, hazra_2024, Walter_disp_readout}, we find $\rm{SNR} \simeq 7.66$. The SNR can be linked to the measurement to the readout fidelity $F$ as \cite{Gambetta2008}
\begin{align} 
    F = 1 - \frac{1}{2} \rm{erfc} \left( \frac{\rm{SNR}}{2 \sqrt{2}} \right). 
\end{align}
Note that this definition implies that $F \in [0.5, 1]$. For $\rm{SNR} \simeq 7.66$, we find $F \simeq 0.9999(4)$, suggesting that the above choice of parameters and threshold comfortably lead to a readout fidelity exceeding $99.99\%$. Of course, the above expression does not account for distortions to the coherent states, transient dynamics of the readout, the $T_1$ limit of the transmon, and other effects. Nevertheless, they provide a useful and simple metric for relating readout performance to the critical photon number. In the main text and \cref{sec:SM stochastic} we go beyond these simple approximations.

\section{Photon number dependent dispersive shift} \label{sec:chi_vs_n}

The dispersive shift is often defined as $\chi_z = (E_{\overline{1,1}} - E_{\overline{1,0}} - E_{\overline{0,1}} + E_{\overline{0,0}}) / 2$, where $E_{\overline{i_t, n_r}}$ denotes the energy of the dressed state $\ket{\overline{i_t, n_r}}$ of the transmon-resonator system, as labeled in \cref{Branch_analysis}. However, this definition of the dispersive shift is based on the low-energy spectrum of the transmon-resonator system and becomes inaccurate even for moderate photon numbers in the resonator \cite{Gambetta:2006, Schuster:2005}.
\begin{figure}
    \centering
    \includegraphics[width=\linewidth]{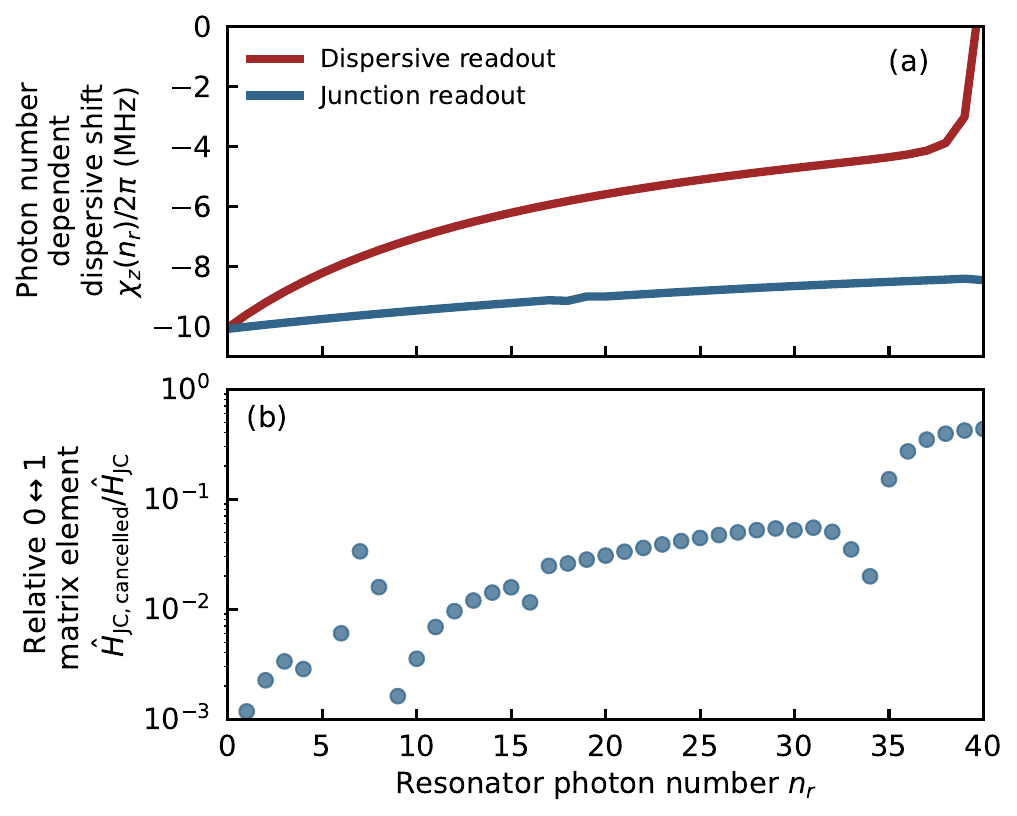}
    \caption{(a) Photon number dependent dispersive shift $\chi_z$ for varying resonator photon numbers $n_r$. For both readout methods, we choose parameters such that $\chi_z(0) / 2 \pi \simeq 10$ MHz. For junction readout, the transmon parameters are the same as in Fig.~4. For dispersive readout, the transmon parameters are the same as in Fig.~3, but the resonator frequency was set to $\omega_r / 2 \pi = 6.246$ GHz, with a coupling strength of $g / 2 \pi = 114$ MHz, to ensure a high critical photon number as shown in Fig.~3. 
    (b) The ratio between the $0 \leftrightarrow 1$ matrix element of the Jaynes-Cummings interaction Hamiltonian with optimal charge-charge coupling $\h{H}_{\mathrm{JC, cancelled}}$ to that without charge-charge coupling $\h{H}_{\mathrm{JC}}$ ,
    at varying resonator photon numbers,  see \cref{sec:chi_vs_n} for further details. 
    }
    \label{fig:chi_vs_n}
\end{figure}

In \cref{fig:chi_vs_n}(a), we compare the magnitude of the dispersive shift at different resonator photon numbers, defined as
\begin{align}
    \chi_z(n_r) = (E_{\overline{1_t, n_r + 1}} - E_{\overline{1_t, n_r}} - E_{\overline{0_t, n_r + 1}} + E_{\overline{0_t, n_r}}) / 2,
\end{align}
for both dispersive readout and junction readout. At zero photons in the resonator, the parameters are taken such that the dispersive shift is $|\chi_z| / 2 \pi \simeq 10$ MHz for both readout methods. For dispersive readout, we find that at high photon numbers the magnitude of the dispersive shift is reduced by more than a half. Conversely, for junction readout the  magnitude of the dispersive shift remains approximately constant. For example, at $n_r = 35$, the dispersive shift is reduced to $\chi_z(n_r = 35) / 2 \pi = -4.37$ MHz for dispersive readout, while $\chi_z(n_r = 35) / 2 \pi = -8.526$ MHz for junction readout. 

We attribute the significant difference between dispersive readout and junction readout to the nature of the coupling interaction. In the dispersive case, exact diagonalization of the Jaynes-Cummings Hamiltonian reveals that, beyond the linear dispersive regime, higher-order contributions to $ \chi_z (n_r)$ scales as $\sim (g/\Delta)^{2n_r}$~\cite{Boissonneault_qubit_readout,Boissonneault_dressed_dephasing,Boissonneault_qubit_dephasing}. Conversely, for the $\cos\hat{\varphi}_r\cos\hat{\varphi}_t$ interaction, these higher-order contributions scale as  $\varphi_\text{zpfr}^{2 n_r}$ \cite{chapple:2024}. For the chosen set of parameters optimized to ensure high $n_{\rm{crit}}$ and $\chi_z$  for both readout methods, we find $\varphi_\text{zpfr}<|g/\Delta|$, and while $g/\Delta$ cannot be made arbitrarily small—otherwise, the leading-order dispersive shift becomes negligible— $\varphi_\text{zpfr}$ can be reduced by decreasing the resonator impedance, without sacrificing the leading-order dispersive shift.

Thus, even when both junction readout and dispersive readout are designed with the same initial dispersive shift $\chi_z(0)$, and both schemes are optimized for high critical photon numbers, junction readout is expected to enable faster measurement, particularly at higher photon numbers.

In addition, \cref{fig:chi_vs_n}(b) shows the ratio between the matrix elements $\vert \langle \overline{0_t, n_r + 1} \vert \h{H}_{\mathrm{JC, cancelled}} \vert \overline{1_t, n_r} \rangle \vert$ and $\vert \langle \overline{0_t, n_r + 1} \vert \h{H}_{\mathrm{JC}} \vert \overline{1_t, n_r} \rangle \vert$ for varying resonator photon numbers. Here, $\h{H}_{\mathrm{JC, cancelled}} = -E_{Jc} \sin{\h{\varphi}_t} \sin{\h{\varphi}_r} + J \h{n}_t \h{n}_r$ with $J$ chosen to satisfy Eq.~(3), and $\h{H}_{\mathrm{JC}} = -E_{Jc} \sin{\h{\varphi}_t} \sin{\h{\varphi}_r}$. We note that even at large resonator photon numbers, the ratio of the matrix elements remains small, showing that the cancellation condition remains effective during readout when the resonator is populated.

\section{Details on readout simulations}
\label{sec:SM stochastic}

In this section, we outline the details of our single-shot readout simulations. For the heterodyne readout simulations of Fig.~4, we use QuTiP's  solver \cite{Qutip_1, Qutip_2} to integrate the stochastic Schr\"{o}dinger equation    
 \cite{Wiseman1993}
\begin{align}
    d \psi(t) &= -i \h{H} \psi(t) dt \nonumber \\ 
    &- \left( \frac{\kappa}{2} \had \ha - \frac{\kappa}{4} [ \langle \h{x} \rangle + i \langle \h{p} \rangle ] \ha + \frac{\kappa}{16} [\langle \h{x} \rangle^2 + \langle \h{p} \rangle^2 ] \right) \psi(t) \nonumber \\
    &+ \sqrt{\frac{\kappa}{2}} \left( \ha - \frac{\langle \h{x} \rangle}{2} \right) \psi(t) \: dW_x \nonumber \\
    &- \sqrt{\frac{\kappa}{2}} \left( i \ha + \frac{\langle \h{p} \rangle}{2} \right) \psi(t) \: dW_p. \label{eqn:SSE}
\end{align}
Here, $\h{H}$ is the sum of Eq.~(1) and the drive term $\h{H}_d = -i \epsilon(t) \cos(\omega_dt)(\ha - \had)$, where $\omega_d$ is the drive frequency and $\epsilon(t)$ is the resonator drive amplitude, with its explicit time dependence provided below. Additionally, $\kappa$ is the resonator decay rate, and $\h{x} = \ha + \had$, and $\h{p} = i (\had - \ha)$ are the measured quadratures. In the above expression, the first and second lines describe the deterministic evolution of the state $\psi(t)$ while the last two terms describe the measurement back action. The two  $dW_i$ are independent stochastic Wiener increments satisfying $\mathbb{E}[dW_i] = 0$ and $\mathbb{E}[dW_i^2] = dt$. Both quadratures, $x$ and $p$, have their own independent Wiener increments, denoted by their respective subscripts.

For each simulated trajectory, we demodulate the signal with $e^{i \omega_d t}$. The demodulated signal for the $x$ and $p$ quadratures are then integrated with the real and imaginary part of the optimal weight function $\langle a_e(t) \rangle - \langle a_g(t) \rangle$, respectively. The results are shown in Fig.~4(a) and (b).
At each integration time we find the optimal discriminator in the IQ-plane which gives us the minimum assignment error defined as $\varepsilon = (P(e \vert g) + P(g \vert e)) / 2$, which is then shown in Fig.~4(c). We found that using \textit{nsubsteps} = $12,000$, with the transmon dimension set to $7$ and resonator dimension set to $75$ showed good convergence. We confirm with branch analysis that this truncation of the Hilbert space does not alter the ionization critical photon number or the dispersive shift.

\begin{figure}
    \centering
    \includegraphics[width=\linewidth]{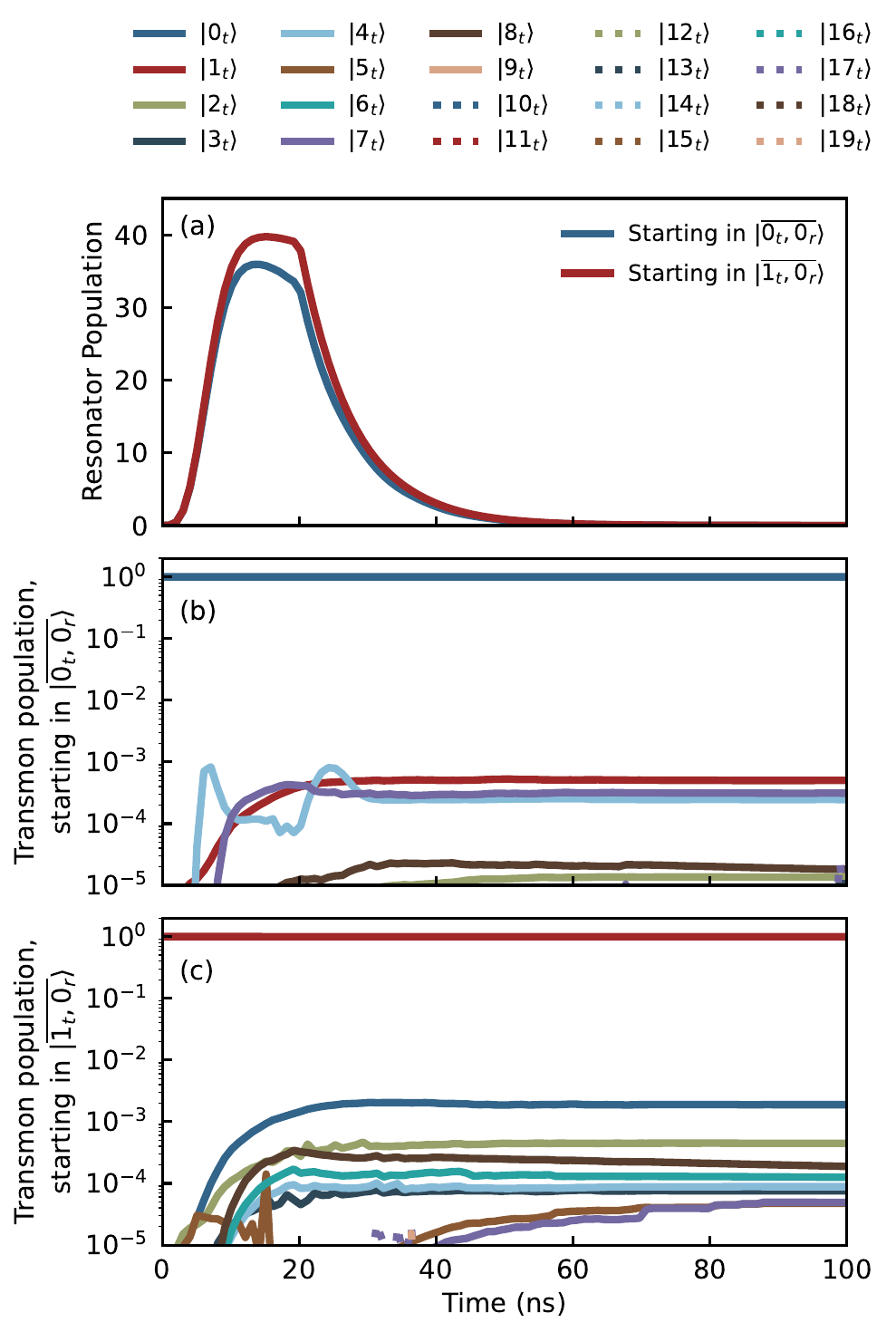}
    \caption{(a) Resonator population as a function of integration time, for when the transmon is prepared in the ground or excited state. (b) and (c) show the population of the first $20$ energy levels of the transmon during the readout simulation, when starting in the ground and excited state, respectively. The circuit parameters are as given in the caption of Fig.~4.
    }
    \label{fig:leakage}
\end{figure}

As mentioned in the main text, we use a two-step pulse in order to fill the resonator with a large number of photons rapidly. The pulse shape is given as in Ref. \cite{chapple:2024},
\begin{align} \label{Two-step-pulse}
    \epsilon(t) = \frac{4 t \sqrt{\Bar{n}}}{\tau^2} e^{-(t / \tau)^2} + \kappa \sqrt{\Bar{n}} (1 - e^{-(t/\tau)^2}),
\end{align}
where $\tau$ and $\kappa$ determine the speed at which the resonator gets populated. Here, we set $\tau$ to be $6$ ns. As we show in \cref{optimal_control_SM}, optimizing the pulse shape with quantum optimal control techniques results in further improvements in performance.

The results from the heterodyne readout simulations are compared to coherent state approximated assignment errors. Assuming the states in the resonator to be coherent states, we can get an expression for the optimal signal-to-noise ratio (SNR) as \cite{Bultink2018}
\begin{align}
    \mathrm{SNR}(t_\mathrm{m}) = \sqrt{2 \kappa \eta \int_0^{t_\mathrm{m}} \vert \alpha_{e}(t) - \alpha_{g}(t) \vert^2 \: dt},
\end{align}
where $\eta$ is the measurement efficiency, $\kappa$ is the resonator decay rate, $t_m$ is the integration time, and $\alpha_{e/g}(t)$ is the resonator field. Using the above formula for the optimal SNR, the assignment error is bounded by  \cite{Gembetta_optimal_readout, swiadek2023enhancing}
\begin{align}
    \varepsilon(t_m) = \frac{1}{2} \mathrm{erfc}\left(\frac{\mathrm{SNR}(t_m)}{2 \sqrt{2}}\right) + \frac{t_m}{2 T_1},
\end{align}
where $\mathrm{erfc}$ is the complementary error function and $T_1$ is the transmon lifetime. The coherent state approximated assignment errors are computed from performing Monte-Carlo simulations using QuTiP's \textit{mcsolve} solver \cite{Qutip_1, Qutip_2} to get the resonator's field $\alpha_{e/g}(t)$. The Hamiltonian and the parameters used for the simulations are the same as what was used for the heterodyne readout simulations. The collapse operator is $\sqrt{\kappa} \ha$ and we use $512$ trajectories for the Monte-Carlo simulations.

\section{Leakage and QNDness} \label{sec:leakage_and_qndness}

\begin{figure}
    \centering
    \includegraphics[width=\linewidth]{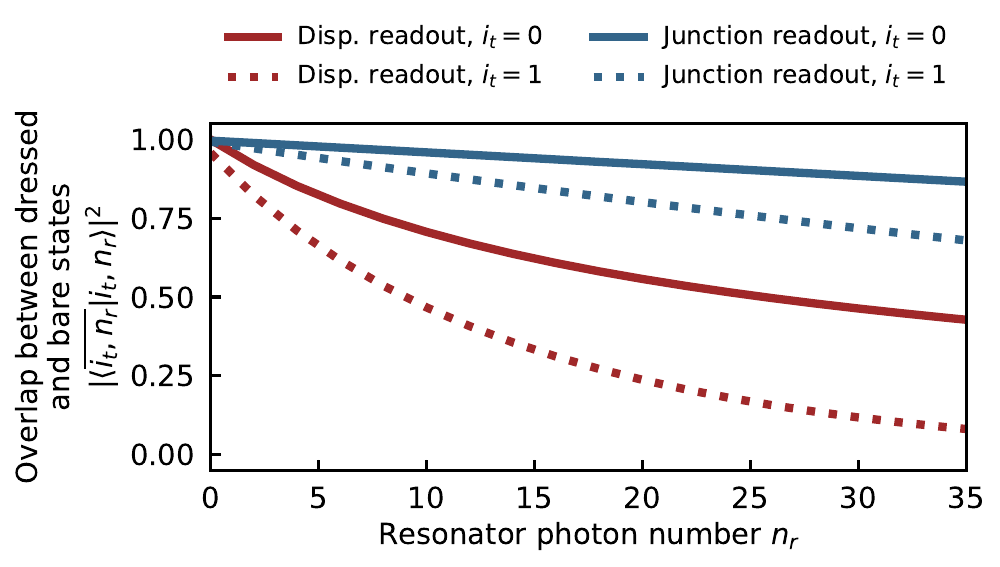}
    \caption{Overlap between the dressed states $\vert \overline{i_t, n_r} \rangle$ and bare states $\vert i_t, n_r \rangle = \vert i_t \rangle \otimes \vert n_r \rangle$ at varying resonator photon numbers, for both junction readout and dispersive readout. The parameters for both readout methods are the same as was described in \cref{fig:chi_vs_n}.}
    \label{fig:bare_state_overlaps}
\end{figure}

In this section, we characterize the leakage rates in our junction readout simulations and provide general comparisons of the expected leakage rates between dispersive and junction readout.
\cref{fig:leakage}(a) shows the resonator population as a function of integration time for the junction readout simulation parameters used in Fig.~4, obtained from the Monte-Carlo readout simulation described in \cref{sec:SM stochastic} by preparing the qubit in the ground state (blue) or excited state (red).

\renewcommand{\arraystretch}{2}
\begin{table*}[t]
    \centering
    \begin{tabular}{c c c c}
    \hline \hline \text { Description } & \text { Analytical formula } & \text { Weight } & \text { Hyperparameters } \\
    \hline \text{Inverse signal-to-noise ratio } & $\left(2 \eta \kappa \int_0^{t_m}\left|\alpha_e(t)-\alpha_g(t)\right|^2 \mathrm{~d} t\right)^{-1/2}$ & $1.0$ & $\kappa / 2 \pi \simeq 20 \: \mathrm{MHz}$ \\
    \text{Maximum pulse amplitude } & $\frac{1}{t_m} \int_0^{t_m} \operatorname{ReLU}\left(\left|\epsilon(t)\right|-\epsilon_{\max }\right) \mathrm{d} t$ & $10^{-4}$ & $\epsilon_{\max } / 2 \pi=700 \: \mathrm{MHz}$ \\
    \text{Maximum resonator population} & $\sum_{i=g, e} \frac{1}{t_m} \int_0^{t_m} \operatorname{ReLU}\left(\operatorname{Tr}\left[\hat{a}^{\dagger} \hat{a} \hat{\rho}_i(t)\right]-\bar{n}_{\text {max }}\right) \mathrm{d} t$ & $10^{-3}$ & $\bar{n}_{\text {max }}=40$ \\
    \text{ Pulse smoothness} & $\frac{1}{t_m} \int_0^{t_m} \left(\partial\epsilon(t)/\partial t \right) \mathrm{d} t$ & $10^{-3}$ & $-$ \\
    \text{Final resonator population} & $\sum_{i=g,e}\operatorname{ReLU}\left(\operatorname{Tr}\left[\hat{a}^{\dagger} \hat{a} \hat{\rho}_i(t_{\rm{total}})\right]\right)$ & $10^2$ & $-$ \\
    \hline \hline
    \end{tabular}
    \caption{Summary of readout cost function used for optimizing the pulse, alongside their analytical expressions, relative weights, and the parameters used. $\operatorname{ReLU}$ denotes the rectified linear unit defined as $\operatorname{ReLU}(x)$ = 0 for $x\leq0$ and $\operatorname{ReLU}(x) = x$ for $x>0$.}
    \label{tab:Cost}
\end{table*}

As previously mentioned, we employ a two-step pulse, stopping the drive at $t = 20$ ns, letting the resonator decay rapidly thereafter with a rate $\kappa/2\pi = 2|\chi_z|/2\pi \simeq 20$ MHz. To quantify the QNDness, we track the population of each transmon level $i_t$ by calculating $P_{i_t}(t) = \mathrm{Tr}\left[ \sum_{n_r} \vert \overline{i_t, n_r} \rangle \langle \overline{i_t, n_r} \vert \h{\rho}(t) \right]$. The transmon populations as a function of time for initial states $\vert \overline{0_t, 0_r} \rangle$ and $\vert \overline{1_t, 0_r} \rangle$ are shown in \cref{fig:leakage}(b) and (c), respectively. From \cref{fig:leakage}(a) we identify the time at which the resonator population drops below $10^{-2}$ and use this to evaluate the population of the ground and first excited states of the transmon to characterize the quantum nondemolition (QND) nature of the readout. For these parameters, the QND fidelity was found to be $99.89\%$ when starting in the ground state and $99.71\%$, when starting in the first excited state. Furthermore, the leakage from the computational states to the second excited state was found to be $0.045\%$ and $0.0014\%$ for the ground and excited state, respectively.

While weak resonances in the transmon-resonator spectrum are often disregarded when analyzing measurement-induced transitions, they are crucial for characterizing leakage rates. The extent to which the transmon leaks depends significantly on the degree of hybridization between the transmon and resonator states~\cite{MIST_2,Dumas2024}. To fully quantify this hybridization, in \cref{fig:bare_state_overlaps}, we plot the overlap between the bare and dressed states of the transmon-resonator system for various resonator photon numbers. We find that for dispersive readout, the overlap decreases rapidly even at moderate photon numbers, particularly when the system starts in the first excited state. In contrast, for junction readout, the overlap exhibits only minimal variation with increasing photon numbers, suggesting that junction readout results in less leakage compared to dispersive readout, as the system's state deviates less from its initial qubit state during the readout process.

\section{Quantum optimal control} \label{optimal_control_SM}

In this section, we present the quantum optimal control (QOC) framework used to obtain the results in Fig.~4(c). In addition to improve the readout fidelity, we also show here that the same QOC method enhances resonator reset, yielding a total readout time (integration plus reset) as short as $t_{\rm{total}} \simeq 75 $ ns with residual photon population $N_r(t_{\rm{total}})\simeq 5 \times 10^{-4}$ with $N_r(t) =  \sum_{i_t,n_r}n_r  \overline{\bra{i_t,n_r}}\hat{\rho}(t) \overline{\ket{i_t,n_r}}$.

These simulations build on Ref.~\cite{gautier2024optimal} and make use of the open-source library dynamiqs \cite{guilmin2025dynamiqs}. To efficiently explore the large optimization parameter space for readout, we employ the evosax library \cite{evosax2022github} together with the Covariance Matrix Adaptation (CMA) evolution strategy. For the reset optimization, we use the optimistix library \cite{optimistix2024} with the BFGS algorithm, ensuring rapid convergence to short reset times.

As discussed in Ref.~\cite{gautier2024optimal}, the readout pulse is divided into $1$ ns time bins, representing the parameters to be optimized, and filtered through a Gaussian envelope with a $294$ MHz bandwidth. Furthermore, the maximum amplitude is limited to $\epsilon_{\textrm{max}} / 2\pi = 700$ MHz whilst ensuring that the photon population does not exceed $\Bar{n}_\textrm{max}=40$ to remain consistent with our previous simulations (see Fig.~4 and \cref{sec:SM stochastic}).

\begin{figure}[t!]
    \centering
    \includegraphics[width=\linewidth]{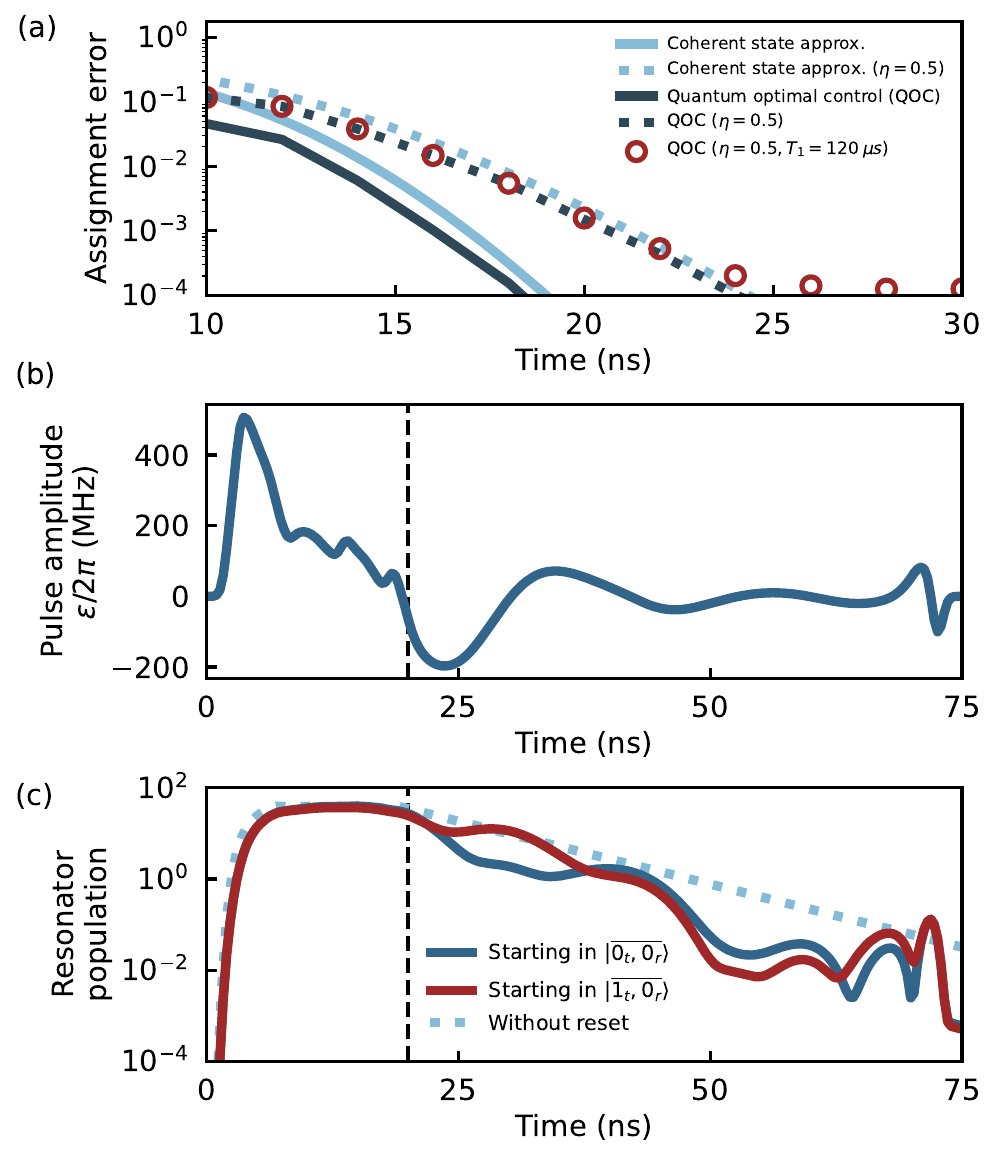}
    \caption{(a) Assignment error obtained from the coherent state approximation (light blue, see Fig.~4(c)) and quantum optimal control (dark blue). Solid lines correspond to the ideal efficiency case of $\eta=1$, while dashed lines correspond to reduced efficiency of $\eta=0.5$, based on state-of-the-art values. Red circles indicate when both finite lifetime ($T_1 = 120~\mu s$) and reduced readout efficiency ($\eta=0.5$) are taken into account. For both methods, the parameters are identical to those in Fig.~4. (b) Optimal pulse shape obtained for a total readout time of $t_{\rm{total}} = 75$ ns including resonator reset, and (c) the corresponding resonator population. Dashed vertical lines indicate $t_m = 20$ ns, the time at which the pulse is optimized to reset the resonator. 
    }
    \label{fig:QOC}
\end{figure}

Focusing first on optimizing the readout fidelity, the cost function consists of weighted sum of 4 components, as summarized in \cref{tab:Cost}: the inverse of the signal-to-noise ratio (SNR), the maximum resonator population, the maximum amplitude, and the pulse smoothness. The optimization is performed for integration times between 10 and 30 ns. Although the pulse shapes vary slightly with integration time, all optimized pulses resemble a two-step pulse and reach the maximum resonator population. Crucially, as shown in \cref{fig:QOC} (a), the optimized pulses outperform the analytical pulse described in \cref{sec:SM stochastic} at all integration times, reaching a readout fidelity of $99.99$\% in $t_m \simeq 18 \: \textrm{ns}$. Even when accounting for a finite $T_1$ of $120 \: \mu s$ and a measurement efficiency of $\eta = 0.5$, the optimal pulse achieves $99.99$\% readout fidelity in $t_m \simeq 28 \: \textrm {ns}$.

To include the reset optimization, we add to the previous cost function a term that penalizes large final resonator population (last row of \cref{tab:Cost}). The previously optimized readout pulses are then re-optimized with this new cost function to accelerate the reset. The resulting readout-and-reset pulse, shown in \cref{fig:QOC}(b), begins by closely resembling the prior optimal readout pulse, while its final segment is reminiscent of a CLEAR pulse \cite{McClure2016}. This pulse still reaches the maximum photon number during readout, yet drives the resonator population back near zero at the end, as shown in \cref{fig:QOC}(c). Importantly, for $\eta = 0.5$ and $T_1 = 120 \: \mu$s, the readout maintains high fidelity—even with resonator reset—exceeding $99.98\%$ at $t_m \simeq 32$ ns, while simultaneously reducing the residual resonator population to $5 \times 10^{-4}$ in $t_{\rm{total}} \simeq 75$ ns. This corresponds to an improvement of $\simeq 4 \: T_{\textrm{res}}$ over passive resonator decay, where $T_{\textrm{res}} = 1/\kappa$ (dashed line in \cref{fig:QOC}(c)). For comparison, using a CLEAR pulse Ref.~\cite{McClure2016} experimentally demonstrated an improvement of $\simeq 2 \: T_{\textrm{res}}$, though this can depend on parameter choices. Finally, it is worth noting that the optimal reset pulse used here inherently takes into account the system's full nonlinearities as captured by Eq.~(1).

\section{Removing the flux loop} \label{sec:fluxloop_SM}

\begin{figure}
    \centering
    \includegraphics[width=\linewidth]{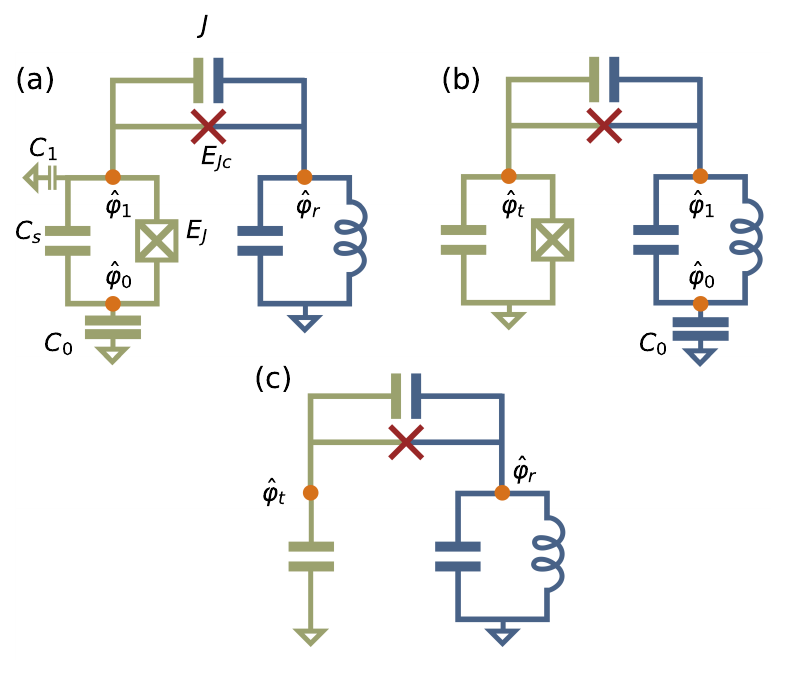}
\caption{Circuit implementations of junction readout without a flux loop.
(a) Floating transmon coupled to a $\lambda/4$ readout resonator.
(b) Grounded transmon coupled to a $\lambda/2$ readout resonator.
(c) Single-junction design in which the transmon nonlinearity is entirely inherited from the coupling junction.
In contrast to configurations (a) and (b), the circuit in (c) does not feature any parasitic low-frequency modes.}
    \label{fig:circuit_appendix}
\end{figure}

The lumped-element circuit of Fig.~1(a) contains a loop through which a flux can be applied. This can be used to tune the qubit but can also lead to dephasing. In this section we discuss three possible designs for implementing junction readout without the flux loop. The circuit designs are shown in \cref{fig:circuit_appendix}. The designs in panels (a) and (b) rely on breaking the flux loop by either floating the transmon or by using a $\lambda/2$ resonator. These approaches result in an extra mode in the circuit whose consequences  we discuss below. In contrast, the design of (c) that breaks the flux loop without introducing an additional mode.

\subsection{Floating transmon design} \label{floating_transmon}

Consider the circuit in \cref{fig:circuit_appendix}(a). The total circuit Hamiltonian can now be written as
\begin{equation} \label{appendix: hamiltonian_floating}
    \hat{H} = \hat{H}_{\text{t,r}} + \hat{H}_{\text{cm}} + \hat
    {H}_{\text{int}},
\end{equation}
where $\hH_{\text{t,r}} $ describes the Hamiltonian of the transmon-resonator system, see Eq.~(1) in the main text. Moreover, $\hH_{\text{cm}}$  and $\hH_{\text{int}}$ represents the Hamiltonian of the additional mode and its interaction with the rest of the system, respectively. Below, we investigate the consequences of this additional mode. For simplicity  we focus on the regime where $C_1\ll C_0$, such that the transmon and the additional mode reduced flux variables are given by  $\hat{\varphi_t} =\hat{\varphi}_1-\hat{\varphi}_0$ and $\hat{\varphi}_{cm} =(C_1/C_0)\hat{\varphi}_1+\hat{\varphi}_0\simeq \hat{\varphi}_0$ \footnote{This simplification also ensures a fair comparison with the grounded case since it maintains similar strong coupling between the transmon and the readout resonator. Alternatively, if $C_0 \simeq C_1$, the charge–charge coupling is reduced by a factor of two, as expected from floating the qubit, and the cross-Kerr interaction is similarly reduced by a factor of four relative to the asymmetric case.}. The conclusion that follow  remain unchanged if the condition $C_1\ll C_0$ is not satisfied.

The presence of this additional mode---often referred to as the ``common mode'' (cm)---in a floating transmon is not unique to junction readout. However, if the transmon is capacitively coupled to the resonator, as in the standard dispersive readout, the common mode is of no consequence.  Indeed, its Hamiltonian consists only of a kinetic charging energy term, i.e., $\hH_{\text{cm}} = 4 E_{C_{cm}} (\hat{n}_{cm} - n_{g_{cm}})^2 $, and it therefore behaves as a free particle. Moreover, for the same reason, the Hamiltonian of the common mode commutes with the charge operator which is the relevant qubit-resonator coupling operator in the case of dispersive readout where the interaction with the common mode reads 
\begin{equation}
    \h{H}_{\text{int,cap.}} = J_{cm,t}\hat{n}_{cm}\hat{n}_{t}+ J_{cm,r}\hat{n}_{cm}\hat{n}_r.
\end{equation}
This interaction is also of no consequence for the transmon and the resonator, i.e., it acts as a renormalization of the effective gate charge acting on the transmon. On the other hand, because of the galvanic coupling between the transmon and the readout resonator in junction readout, the common mode acquires a potential-like term, i.e., $\hH_{\text{cm}} = 4 E_{C_{cm}} (\hat{n}_{cm} - n_{g_{cm}})^2 -E_{J_c}\cos\hat{\varphi}_{cm} $, and non-trivial coupling to both the transmon and readout resonator modes,
\begin{align}
\begin{split} 
    &\hH_{\text{int,jr}} = J_{cm,t}\hat{n}_{cm}\hat{n}_t+ J_{cm,r}\hat{n}_{cm}\hat{n}_r\\&-E_{J_c}\left[\cos(\hat{\varphi}_{cm}+\hat{\varphi}_t-\hat{\varphi}_r)-\cos(\hat{\varphi}_{cm})-\cos(\hat{\varphi}_t-\hat{\varphi}_r) \right].
    \end{split}
\end{align}
The last two cosine terms in the second line are present to avoid double-counting contributions already included in $\hH_{\mathrm{t,r}}$ and $\hH_{\text{cm}}$. 
This coupling can result in new multiphoton resonances,  
something which  can lower the ionization critical photon number for the transmon mode.

\begin{figure}
    \centering
    \includegraphics[width=\linewidth]{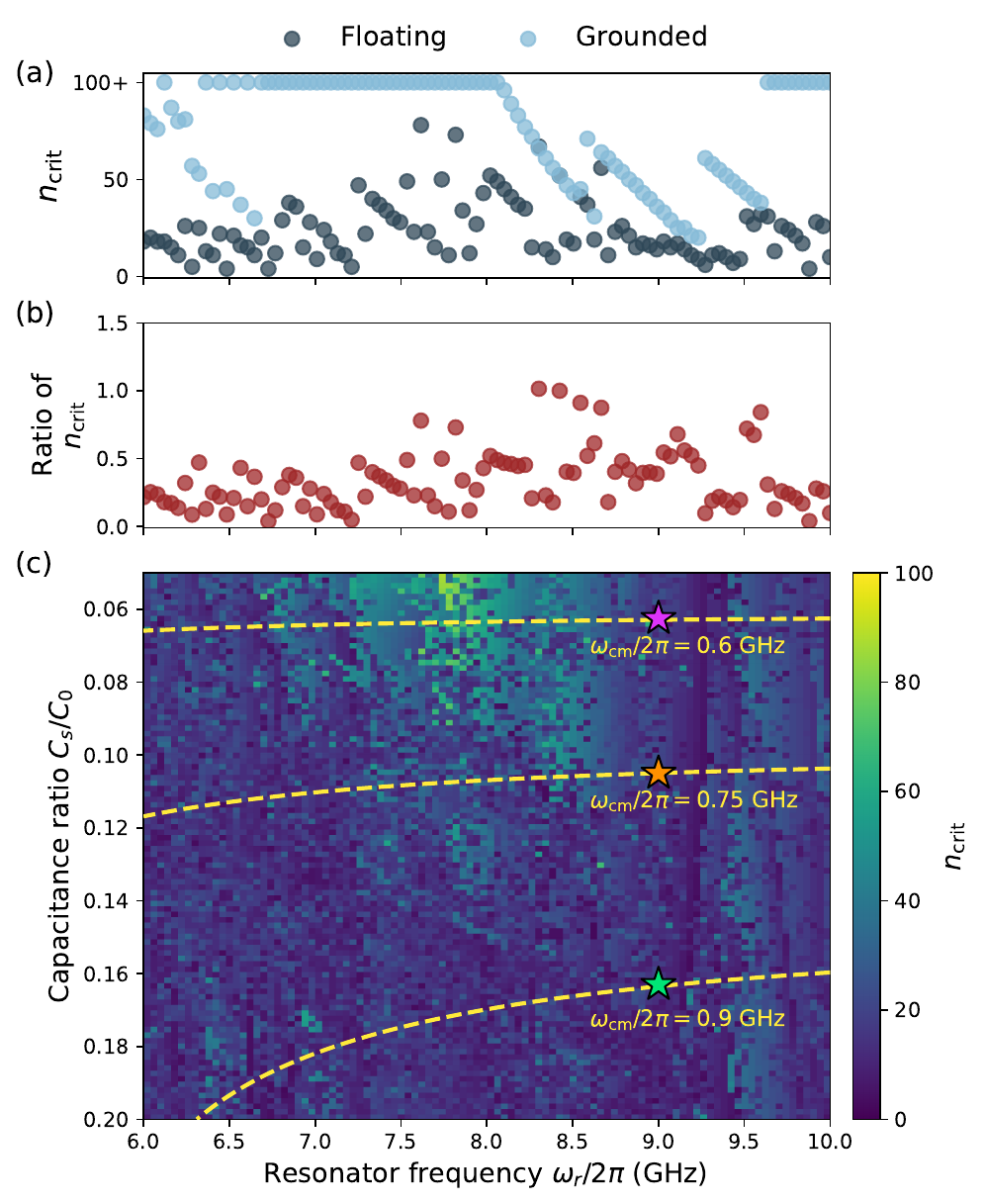}
    \caption{
    (a) Critical photon number, and (b) the ratio between the critical photon numbers for floating and grounded transmon designs, for various resonator frequencies $\omega_r$. Here the ratio $C_s / C_0 = 0.1$. (c) Critical photon number of the floating transmon case for varying $C_s/C_0$ ratios and resonator frequencies $\omega_r$. The contours of several common mode frequencies are shown in dashed yellow lines. The coupling strength $J_{cm,t}$ between the common mode and the transmon at the purple, orange and green stars are $J_{cm.t} / 2\pi \simeq -23.36, \: -41.32$, and $-59.29$ MHz, respectively. The parameters used here are the same as in \cref{sec:extract_purcell}.
    }
    \label{fig:floating_vs_grounded}
\end{figure}

\cref{fig:floating_vs_grounded}(c) shows the critical photon number as a function of the resonator frequency $\omega_r$ and the ratio $C_s/C_0$ obtained for the coupled transmon, common mode, and resonator system using branch analysis. In this analysis, the common mode is assumed to initially be in its ground state. Here, $C_s$ is the transmon shunt capacitance and $C_0$ is the capacitance to ground, see \cref{fig:circuit_appendix}(a), both of which can be chosen at fabrication time. 
We keep the shunt capacitor $C_s$ of the transmon fixed such that the transmon charging energy remains around $E_{C_t}/2\pi \simeq 300$ MHz, similar to the grounded transmon case studied in the main text. All other parameters are identical to those in \cref{sec:extract_purcell}, with common-mode transmon coupling $J_{cm,t} \simeq -J C_r/C_0$ which depends on the ground capacitance $C_0$.
\cref{fig:floating_vs_grounded}(a) compares the critical photon number in the floating design with $C_s/C_0=0.1$ to the grounded case, and \cref{fig:floating_vs_grounded}(b) shows the corresponding ratio. We observe that in most instances floating the transmon lowers the critical photon number compared to the grounded case. This is similar to spectator-induced state transitions recently studied in strongly driven multimode systems \cite{singh:2025, connolly2025,dai2025,Hoyau:2024,benhayounekhadraoui2025}.

Interestingly, from \cref{fig:floating_vs_grounded}(c) we find that decreasing the ratio $C_s/C_0$ increases the critical photon number. This is likely due to the common-mode frequency $\omega_{cm}$ being pushed downward, along with reductions in the coupling strengths $J_{cm,r} \simeq J (C_s/C_0)$ and $J_{cm,t}$. However, reducing $C_s/C_0$ even more can push the common mode into a frequency range where it becomes thermally populated, which can lead to additional dephasing of the transmon mode \cite{Bertet:2005,Gambetta:2006,Clerk:2007,Sears:2012}.

Importantly, we emphasize that the challenges introduced by floating the qubits are not inherent to the junction readout circuit itself but rather a general consequence of using any coupling other than capacitive coupling. Indeed, with inductively coupled readout schemes--or inductive or junction-based two-qubit couplers--this extra mode acquires a potential-like term, leading to similar considerations, see for instance Refs.~\cite{Ma2024, chakraborty2025}.

\subsection{ Using a $\lambda/2$ readout resonator}

\begin{figure}
    \centering
    \includegraphics[width=\linewidth]{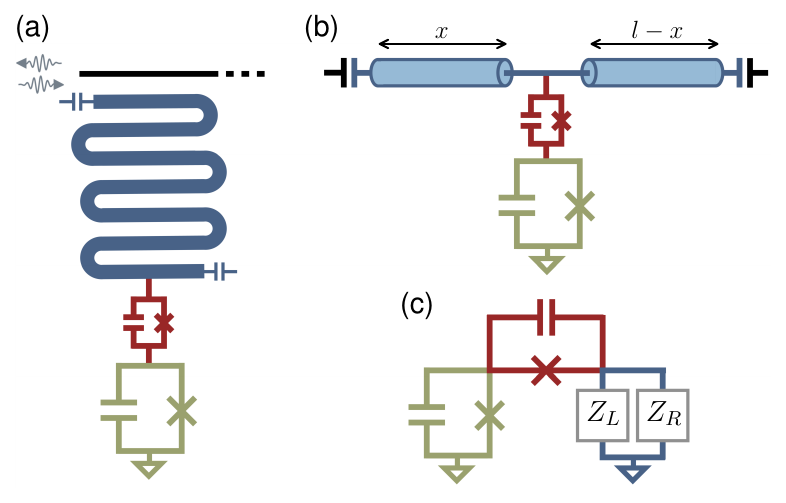}
    \caption{ (a) Circuit implementation where the coupling junction is located at an arbitrary position $x$ along a $l=\lambda/2$ readout resonator. This setup can be modeled as the junction connected to two transmission line segments of length $x$ and $l - x$, respectively as shown in (b).
(c) Equivalent circuit-model in which the junction is coupled to an effective impedance given by the parallel combination of the input impedances of the left and right transmission line segments, denoted $Z_L(\omega)$ and $Z_R(\omega)$.}
    \label{fig:lambda2}
\end{figure}

An alternative way to eliminate the flux loop is to use a $\lambda/2$ readout resonator, as illustrated in \cref{fig:lambda2}(a). The circuit again maps to a floating configuration, but now it is the LC resonator rather than the transmon that is floated, see \cref{fig:circuit_appendix}(b).
To show this, consider the input impedance of the transmission-line (TL) resonator as seen by the coupling junction. For the moment, we consider the case where the coupling junction is located at the far end of TL resonator, i.e., at distance $l=\lambda/2$ from the open end. Below, we analyze the general case of coupling at arbitrary position inside the TL. 

The input impedance seen by the coupling junction is given by \cite{Pozar2011MicrowaveEngineering}
  \begin{equation}
      Z_{\mathrm{in}} (\omega)= -iZ_0\tan^{-1}(\omega l/v),
  \end{equation}
where $Z_0$ and $v$ are the characteristic impedance and phase velocity of the TL, respectively. The relevance of $Z_{\mathrm{in}}(\omega)$ is that its poles identify the resonances of the readout circuitry, while the associated residues set each mode’s impedance—and thus its coupling strength to the qubit—and at the same time accounting for the distributed nature of the circuit~\cite{Nigg2012, Solgun2014, Solgun2015, Solgun2019, Labarca2024}.
Using the fraction expansion 
\begin{equation}
    \tan^{-1}(z) = \frac{1}{z}+\sum_{n=1}^{+\infty}\frac{2z}{z^2-n^2\pi^2}, \: \: z \in \mathbb{C}\setminus \{n \pi,\: n \in \mathbb{Z}\},
\end{equation}
yields the Foster decomposition
\begin{equation} \label{Foster_lambda/2}
    Z_{\mathrm{in}}(\omega) = \frac{1}{i\omega C_0} + \sum_{n=1}^{\infty} \frac{i\omega C_r^{-1}}{\omega_n^2-\omega^2},
\end{equation}
where $\omega_n = n\pi v/l= n \omega_r$, and $C_r = \pi/(2Z_0\omega_r )$. The fundamental mode  of a $\lambda/2$  resonator thus has an impedance $Z_r = 2Z_0/\pi <Z_0$. Crucially, \cref{Foster_lambda/2}  shows that a $\lambda/2$ TL is equivalent to a collection of harmonic oscillators (second term) in series with a ground capacitance $C_0= 2C_r$ (first term). Fig.~1(b)  corresponds to keeping only the fundamental resonance $\omega_r$, i.e., the term $n=1$ in the sum. This approximation holds as long as the operating frequency obeys $\omega \ll 2 \omega_r$.
 
Importantly, the ground capacitance $C_0$ leads to the same considerations discussed in \cref{floating_transmon}: it introduces a low-frequency degree of freedom that couples to both the transmon and the readout mode. Indeed, defining $\hat{\varphi}_r = \hat{\varphi}_1-\hat{\varphi}_0 $, and $\hat{\varphi}_{cm} = -\hat{\varphi}_0$, the circuit Hamiltonian of \cref{fig:circuit_appendix}(b) takes the same form as \cref{appendix: hamiltonian_floating}, with circuit parameters given by  $E_{C_{cm}} = e^2/(2C_0)= E_{C_{r}}/2$, $J_{cm,t}\simeq J/2$, $J_{cm,r}\simeq -J(C_s/2C_r)$. Consequently, the critical photon number for measurement-induced state transitions displays the same features as in \cref{fig:floating_vs_grounded}.

We now consider the more general case where the coupling junction is located at an arbitrary position $x$ along the $l=\lambda/2$ resonator as shown in \cref{fig:lambda2}(a) and (b). Below, we show that the circuit can still be mapped to a floating configuration, as in \cref{fig:circuit_appendix} (b), but with the key difference that the mode impedance $Z_r$ now depends on the coupling position $x$.

In this configuration, the coupling junction is effectively connected to two open-ended transmission-line segments of lengths $x$ and $l-x$, respectively, as illustrated in \cref{fig:lambda2}(b). Denoting their input impedances as $Z_L(\omega)$ and $Z_R(\omega)$, the coupling junction sees a total impedance given by the parallel combination of the two, that is,
\begin{equation} \label{input_impedance_inside}
Z_{\mathrm{in}}(\omega )= \frac{Z_L(\omega)Z_R(\omega)}{Z_L(\omega)+Z_R(\omega)},
\end{equation}
see \cref{fig:lambda2}(c). Here,  $Z_L(\omega) =-iZ_0\tan^{-1}(\omega(l-x)/v)$ and $Z_R(\omega)=-iZ_0\tan^{-1}{(\omega x/v)}$ with $0<x<l$. We assume a homogeneous TL, meaning both segments have the same characteristic impedance $Z_0$ and phase velocity $v$. 

As before, we expand $Z_{\mathrm{in}}$ in its canonical Foster form,
\begin{equation} \label{foster_couplinginside}
        Z_{\mathrm{in}}(\omega) = \frac{A_0}{i\omega} + \sum_{n=1}^{\infty} \frac{i\omega A_n}{\omega_n^2-\omega^2},
\end{equation}
to extract the relevant quantities. The residue of the zero-frequency pole,
\begin{equation}
 A_0(x) =\lim_{\omega \to 0} [i\omega Z_{\mathrm{in}} (\omega)],
\end{equation} determines the effective ground capacitance $C_0$, while the impedance of the readout mode is set by the residue of the first finite pole,
\begin{equation}
    A_1(x)= \lim_{\omega\to \omega_1}[2i(\omega-\omega_1)Z_{\mathrm{in}}(\omega)].
\end{equation}
Using the analytical expressions for $Z_L(\omega) $ and $Z_R(\omega)$, we find that $A_0(x) = Z_0v/l=C_0^{-1}$, which is independent of the coupling position and matches the value obtained when the coupling is at the end of the resonator.  Moreover, from \cref{input_impedance_inside}, the poles of $Z_{\mathrm{in}}(\omega)$ fall into three categories: (i) poles of $Z_L$ located at $n\pi v/x $, $n \in \mathbb{N}_+$ (ii) poles of $Z_R$ at $n\pi v/(l-x) $, with $n \in \mathbb{N}_+$ and (iii) frequencies satisfying the current conservation $Z_L(\omega)=-Z_R(\omega)$ which occur at $n\pi v/l=n\omega_r $, $n \in \mathbb{N}_+$. Thus, the lowest pole corresponds to $\omega_r$, as expected. A straightforward calculation yields the position-dependent impedance of the readout mode,
 \begin{equation}
     Z_r(x)  = \frac{2Z_0}{\pi} \frac{1}{\cos^2(\frac{\pi(l-x)}{l})},
 \end{equation}
with $Z_r(x)=A_1(x)/\omega_r$. This expression shows that the impedance is minimized at the end of the resonator, where $Z_r (x=l)= 2Z_0/\pi<Z_0$, and diverges at the center $x=l/2$, as expected.

For completeness, we briefly comment on the case of a $\lambda/4$ readout resonator, where one end is now shorted. In this configuration, there is no zero-frequency pole, i.e., $A_0=0$, reflecting the absence of a ground capacitance. Physically, this corresponds to the lack of a floating island and thus the absence of a low-frequency mode when the resonator is coupled galvanically to the qubit. Of course, this comes at the cost of introducing a flux loop in the presence of both $E_J$ and $E_{J_c}$, see the next section for a single-junction design. Moreover, the position-dependent impedance of the readout mode is
$Z_r(x) = (4Z_0/\pi)\cos^2[\pi(l-x)/(2l)]$. Unlike the $\lambda/2$ case, the mode impedance is maximal when the coupling junction is placed at the end $x=l$, taking the value $Z_r (x=l) = 4Z_0/\pi > Z_0$. Importantly, $Z_r$ can be reduced by moving the coupling point away from the end.

\subsection{Single junction design}

\begin{figure}
    \centering
    \includegraphics[width=\linewidth]{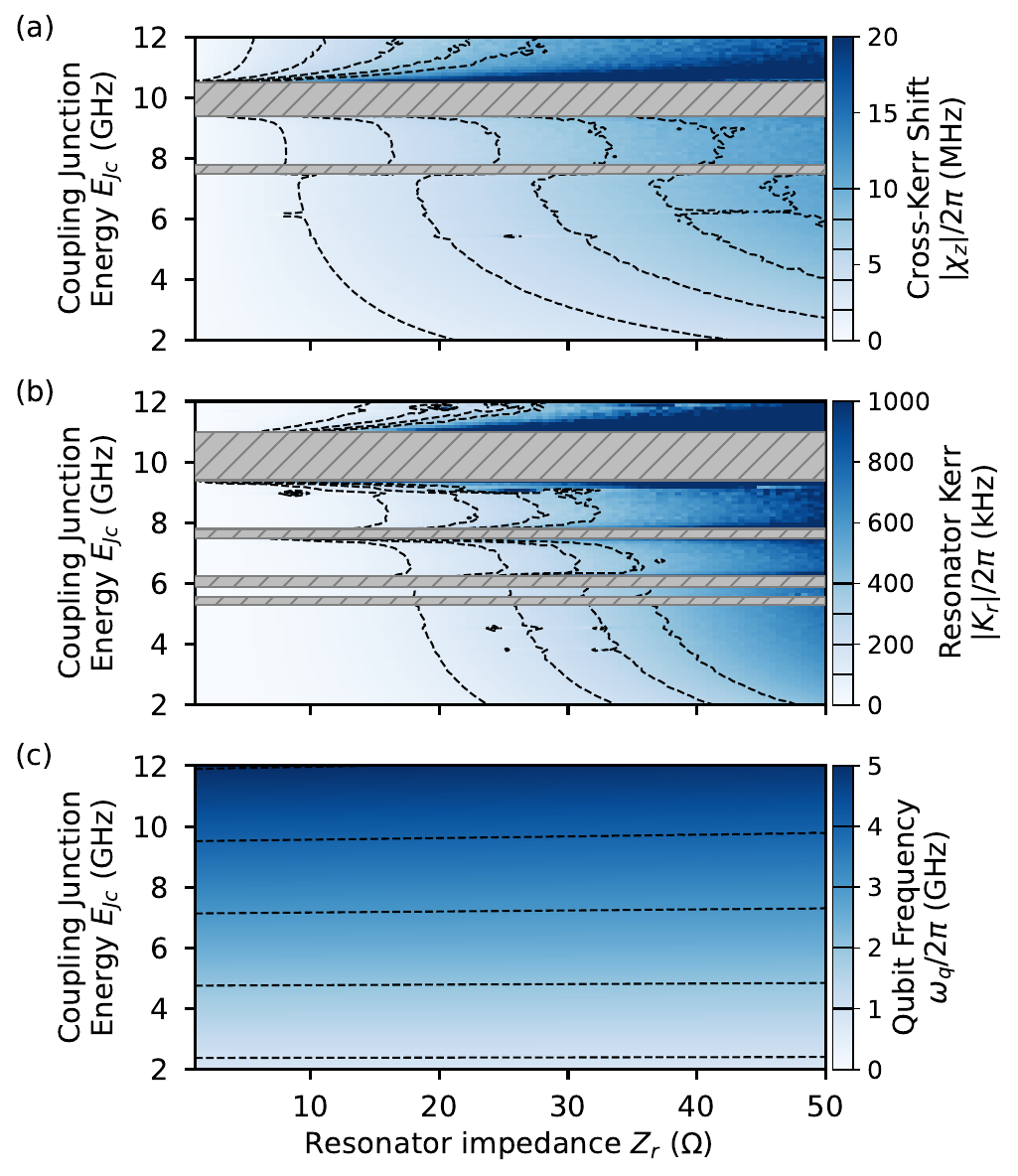}
    \caption{(a) Cross-Kerr coupling strength, (b) resonator self-Kerr, and (c) qubit frequency for various resonator impedances $Z_r$ and coupling junction size $E_{Jc}$, for the single junction design. The resonator frequency was set to $\omega_r/2\pi = 11$ GHz. Grayed out region corresponds to where the dispersive approximation is invalid.}
    \label{fig:single_junction_design_param}
\end{figure}

The circuit of \cref{fig:circuit_appendix}(c) removes the transmon junction. In this case, the transmon’s Josephson energy is set entirely by the coupling junction $E_{J_c}$. Crucially, in this configuration there are no flux loop and no extra modes, and the circuit is reminiscent of the arm qubit proposed in Ref.~\cite{kline:2025}. \cref{fig:single_junction_design_param} shows the cross-Kerr, resonator self-Kerr, and qubit frequencies as functions of the coupling junction energy $E_{J_c}$ and the resonator impedance $Z_r$. As with the circuit of Fig.~1(a), large cross-Kerr with minimal self-Kerr is readily possible. Although $E_{J_c}$ now sets both the coupling strength and the qubit frequency, thereby potentially reducing parameter flexibility, we find this is not limiting, particularly for $\omega_q/2\pi \lesssim 5~\text{GHz}$.

We now explore two regimes of this design: one where $E_{Jc}$ is chosen such that the qubit frequency is in the typical range of $\omega_q / 2\pi = 4 \sim 6$ GHz, and a second where the qubit frequency is below $\omega_q /2\pi = 2$ GHz, while the impedance was set to $Z_r = 18 \: \Omega$ in both cases. 

\begin{figure}
    \centering
    \includegraphics[width=\linewidth]{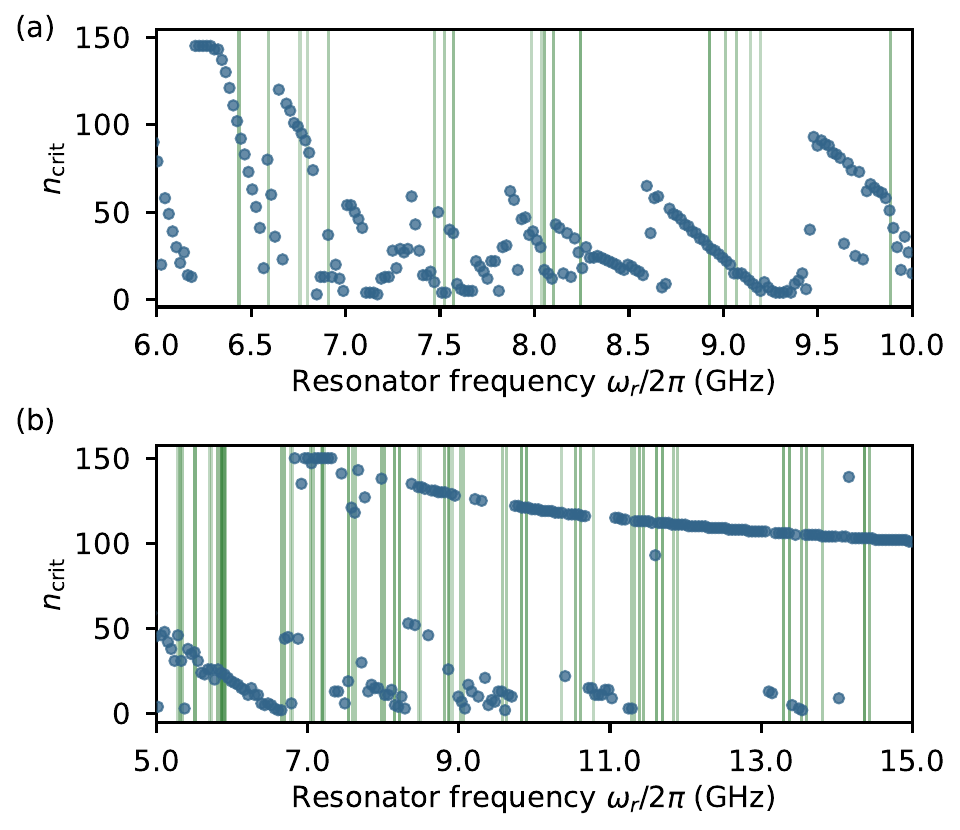}
    \caption{Gate-charge averaged critical photon number for the single-junction design, shown for (a) $\omega_q/2\pi \simeq 4.63$ GHz and (b) $\omega_q/2\pi \simeq 1.65$ 
    GHz, as a function of the readout resonator frequency. The resonator impedance was set to $Z_r = 18 \: \Omega$ for both cases. Vertical lines indicate multiphoton resonances up to a $4$-photon process, with darker lines corresponding to fewer-photon processes. Note the difference in the x-axis ranges between panels (a) and (b).
    }
    \label{fig:single_junction_design_ncrit}
\end{figure}

In the first case, we choose $E_{Jc}/2\pi = 11$ GHz, with $E_C / 2\pi = 275$ MHz, corresponding to a qubit frequency $\omega_q/2\pi = 4.63$ GHz.  \Cref{fig:single_junction_design_ncrit}(a) shows the critical photon number versus resonator frequency obtained for these parameters. Vertical lines mark the locations of multiphoton (up to $4$ photon) resonances that couple the transmon’s ground and first-excited states to higher levels. We find that the critical photon number is large for several ranges of resonator frequencies. With suitable optimization, this single-junction design performs comparably to a standard transmon over the typical qubit-frequency range.

We now focus on the alternative case where the parameters are chosen such that $\omega_q / 2\pi \lesssim 2$ GHz. For our simulations, we choose $E_{Jc} / 2\pi = 4$ GHz and $E_C/2\pi = 100$ MHz, which ensures a large $E_J/E_C$ ratio whilst maintaining an anharmonicity of $\alpha / 2\pi \sim 111$ MHz. Due to the lower qubit frequency, all higher transitions (e.g., $\omega_{03}$ and $\omega_{14}$) also lie at relatively low frequencies. Consequently, numerous resonances occur in their vicinity, leading to low critical photon numbers, see \cref{fig:single_junction_design_ncrit}(b). However, beyond a resonator frequency of approximately $\omega_r/2\pi = 7$ GHz, the critical photon number increases sharply. This behavior mirrors the high-frequency readout of transmons recently reported in Refs.~\cite{connolly2025,kurilovich:2025}, where the resonator frequency was placed an order of magnitude or more above the qubit frequency. Such a large detuning reduces the likelihood of accidental frequency collisions and thereby suppresses multiphoton transitions. Crucially, Refs.~\cite{kurilovich:2025, connolly2025} rely on a charge–charge coupling based dispersive readout, so the large qubit–resonator detuning necessarily comes at the cost of a small dispersive shift. In principle, one could compensate the large detuning by increasing the transmon–resonator coupling capacitance.  In practice, however, the dispersive shift remains small as a larger capacitive coupling results in a lower anharmonicity which is not favorable.

These limitations can be mitigated with the single-junction design of \cref{fig:circuit_appendix}(c) since there is no trade off between the cross-Kerr coupling strength and detuning. In addition, the anharmonicity is no longer restricted by a large capacitive coupling, though it may still be constrained by the $E_{J_c}/E_C$ ratio.

In summary, the single junction design offers similar performance to the standard junction readout design without the inherent flux loop or additional modes in the system. Furthermore, we have shown that high frequency readout of transmons with junction readout may be attractive for high QND readout without the limitations found in the case of dispersive readout.

\end{document}